\documentclass[12pt,preprint]{aastex}
\usepackage{apjfonts}
\usepackage{amssymb, natbib, amsmath, epsfig, epsf, lscape, rotating}
\usepackage{graphicx}
\usepackage[colorlinks,linkcolor=blue,filecolor=blue,urlcolor=blue,
citecolor=blue]{hyperref}
\makeatletter

\newcommand{\Rmnum}[1]{\expandafter\@slowromancap\romannumeral #1@}
\makeatother

\newcommand{\angstrom}{\text{\normalfont\AA}}
\newcommand{\aox}{\ensuremath{\alpha_{\rm{OX}}}}

\newcommand{\bha}{H\ensuremath{\alpha ^{\rm B}}}
\newcommand{\bhb}{H\ensuremath{\beta  ^{\rm B}}}

\newcommand{\civ}{C\,{\footnotesize IV}}
\newcommand{\chisq}{\ensuremath{\chi^2}}

\newcommand{\flux}{erg~s$^{-1}$~cm$^{-2}$}

\newcommand{\fwbha}{\ensuremath{\mathrm{FWHM}_{\rm \bha}}}

\newcommand{\feii}{{\rm Fe\,{\footnotesize II}}}

\newcommand{\ha}{H\ensuremath{\alpha}}
\newcommand{\hb}{H\ensuremath{\beta}}

\newcommand{\hii}{H\,{\footnotesize II}}

\newcommand{\kms}{\ensuremath{\mathrm{km~s^{-1}}}}

\newcommand{\lum}{erg~s\ensuremath{^{-1}}}
\newcommand{\lbol}{\ensuremath{L\mathrm{_{bol}}}}

\newcommand{\ledd}{\ensuremath{L\mathrm{_{Edd}}}}
\newcommand{\lratio}{\ensuremath{\lbol/\ledd}}

\newcommand{\lbha}{\ensuremath{L_{\rm \bha}}}

\newcommand{\lfive}{\ensuremath{\lambda L_{\lambda}}(5100~\AA)}

\newcommand{\llambda}{\lbol$/$\ledd}

\newcommand{\msun}{\ensuremath{M_{\odot}}}
\newcommand{\mbh}{\ensuremath{M_\mathrm{BH}}}

\newcommand{\nii}{[N\,{\footnotesize II}]}

\newcommand{\nh}{\ensuremath{N_{\rm H}}}

\newcommand{\oi}{[O\,{\footnotesize I}]}
\newcommand{\oiii}{[O\,{\footnotesize III}]}

\newcommand{\rosat}{\emph{ROSAT}}

\newcommand{\rkel}{{\emph R}}

\newcommand{\sii}{[S\,{\footnotesize II}]}

\newcommand{\snr}{\rm S$/$N}

\newcommand{\whz}{W~Hz$^{-1}$}

\def\lax{{$\mathrel{\hbox{\rlap{\hbox{\lower4pt\hbox{$\sim$}}}\hbox{$<$}}}$}}
\def\gax{{$\mathrel{\hbox{\rlap{\hbox{\lower4pt\hbox{$\sim$}}}\hbox{$>$}}}$}}
\def\chandra{{\it Chandra}}

\def\newton{{\it XMM-Newton}}

\shorttitle{Seyfert 1 Nuclei with Low-Mass Black Holes} %
\shortauthors{He-Yang~Liu et~al.}

\begin{document}

\title{A UNIFORMLY SELECTED SAMPLE OF LOW-MASS BLACK HOLES 
IN SEYFERT~1 GALAXIES. \Rmnum{2}. THE SDSS DR7 SAMPLE}

\author{
He-Yang~Liu\altaffilmark{1,2,4}, Weimin~Yuan\altaffilmark{1,2},
 Xiao-Bo~Dong\altaffilmark{3},  
Hongyan~Zhou\altaffilmark{4,5} and Wen-Juan~Liu\altaffilmark{3}
}
\altaffiltext{1}{Key Laboratory of Space Astronomy and Technology, National 
Astronomical Observatories, Chinese Academy of Sciences, 20A Datun Road, 
Chaoyang District, Beijing, 100012, China; \mbox{wmy@nao.cas.cn}, 
\mbox{liuheyang@nao.cas.cn} }
\altaffiltext{2}{School of Astronomy and Space Sciences, University of Chinese 
Academy of Sciences, 19A Yuquan Road, Beijing, 100049, China}
\altaffiltext{3}{Yunnan Observatories, Chinese Academy of Sciences, Kunming, Yunnan 650011, China;
Key Laboratory for the Structure and Evolution of Celestial Objects, Chinese Academy of Sciences,
Kunming, Yunnan, 650011, China;\   \mbox{xbdong@ynao.ac.cn} }
\altaffiltext{4}{Polar Research Institute of China, 451 Jinqiao Road, Shanghai 200136, China}
\altaffiltext{5}{The University of Sciences and Technology of China, Chinese 
Academy of Sciences, Hefei, Anhui, 230026, China} 
 

\begin{abstract}
A new sample of 204 low-mass black holes (LMBHs) in active galactic nuclei 
(AGNs) is presented with black hole masses in the range of $(1-20) \times 
10^5$\,\msun. The AGNs are selected from a systematic search among galaxies 
in the seventh Data Release (DR~7) of the Sloan Digital Sky Survey (SDSS), by careful 
analyses of their optical spectra and precise measurement of spectral parameters.
Combining them with our previous sample selected from the SDSS DR~4 
makes it the largest LMBH sample so far, totaling over 
500 objects. Some of the statistical properties of the combined LMBH AGN 
sample are briefly discussed, in the context of exploring the low-mass 
end of the AGN population. Their X-ray luminosities follow the extension 
of the previously known correlation with the \oiii\ luminosity. The effective 
optical-to-X-ray  spectral indices \aox, albeit with a large scatter, are broadly 
consistent with the extension of the relation with the near-UV luminosity 
$L_{\rm 2500 \angstrom}$.  Interestingly, a correlation of \aox\ with black 
hole mass is also found in the sense that \aox\ is statistically flatter (stronger 
X-ray relative to optical) for lower black hole mass. Only 26 objects, mostly 
radio loud, were detected in radio at 20\,cm in the FIRST survey, giving a 
radio loud fraction of 4\%. The host galaxies of  LMBHs have stellar 
masses in the range of $10^{8.8}-10^{12.4}$~\msun\ and optical colors typical 
of Sbc spirals. They are dominated by young stellar populations which seem 
to have undergone a continuous star formation history.

\end{abstract}
 

\setcounter{footnote}{0}
\setcounter{section}{0}


\section{INTRODUCTION}

Mounting evidence has been accumulated suggesting that many massive galaxies 
harbor supermassive black holes (SMBHs) with masses ranging from
10$^6$\,\msun\ to\,10$^9$\,\msun\ at their centers in the present universe 
\citep[e.g.,][]{richstone98,kormendy13}. The mass of SMBHs correlates tightly 
with the parameters of the host bulges such as mass, luminosity and velocity 
dispersion, which hints at a picture of co-evolution of SMBHs and their host 
galaxies \citep[e.g.,][]{magorrian98,ferrarese00,gebhardt00a,gultekin09}. How 
black holes (BHs) have formed across the cosmic time remains far from 
comprehension. It is commonly believed that SMBHs have grown from seed 
black holes through mainly merger-induced accretion \citep {volonteri10, 
greene12}. Secular processes also play a role in BH growth, especially 
for BHs located in low-redshift galaxies \citep{jiangyf11, greene08, conselice14}. 
However, even less is known regarding the properties of seed BHs such as their 
origins, initial masses and environments, simply because direct observations 
of seed BHs in the early universe are not feasible with current facilities. As an 
indirect approach, low-mass black holes (LMBHs) with masses of thousands to 
hundreds of thousands of solar masses, residing at the center of galaxies, 
may help give insight into the formation and evolution of the first seed black 
holes. Such BHs are also termed intermediate-mass black holes (IMBHs) 
in the literature. We refer to them as LMBHs to avoid confusion from Ultra-luminous 
X-ray sources (ULXs), which are off-nucleus point-like sources and some may be 
powered by IMBHs \citep{kaaret17}. LMBHs found in nearby dwarf stellar systems 
in the present universe seem to have formed early, but have not fully grown into 
SMBHs \citep{greene12}. They may thus trace the seed BHs or the early phase 
of their growing. In addition, the study of the occupation fraction of  the 
present-day LMBHs can help discriminate various seed BH models, such as light 
seeds as the end-products of Population III stars or heavy seeds formed from 
the direct collapse of halos at high redshifts \citep[e.g.,][]{volonteri08,volonteri09,
lodato06}. Interestingly, mergers of binary BHs with masses of the order of $10^5$ 
\msun\,can produce gravitational-wave signatures \citep[e.g.,][]{hughes02} that 
are the primary targets of LISA\footnote {Laser Interferometer Space Antenna 
(LISA) is a space-based gravitational wave observatory led by the European 
Space Agency (ESA), designed to detect and accurately measure gravitational 
waves. The new LISA mission (based on the 2017 L3 competition) is a collaboration 
of ESA and NASA.} in the future. The detection of gravitational waves in these 
frequency regimes has become more promising after the prototypical event 
GW150914, the first-ever gravitational-wave source detected by the Laser 
Interferometer Gravitational-wave Observatory \citep[LIGO; e.g.,][]{abbott16}.

The most direct and secure way to detect BHs at the center of galaxies is to 
seek the effect of their gravitational influence on the spatially resolved dynamics 
of closely surrounding gases and/or stars \citep[e.g.,][]{barth01,ghez08}. However, 
for LMBHs, this method is infeasible for distant galaxies since the BH gravitational 
sphere of influence can only be spatially resolved within the Local Group with 
the existing facilities. A common practice is to search for accretion-powered 
radiative signatures of active galactic nuclei (AGNs) hosting such LMBHs. For 
AGNs exhibiting broad emission lines in their optical spectra, their BH masses 
can be estimated using \mbh\,=\,$fR\upsilon^2/G$, assuming that the broad line 
region (BLR) system is virialized and individual clouds are moving in Keplerian 
orbits. $G$ is the gravitational constant. The velocity dispersion $\upsilon$\ 
can be measured from the widths of the broad emission lines.
The BLR radius $R$ is estimated from the luminosity of AGN continuum emission 
using the radius-luminosity relation derived from reverberation mapping studies of 
AGNs \citep[e.g.,][]{kaspi05,bentz06,bentz09a,bentz13}. $f$ is a scaling factor 
of the order of unity, depending on the distribution and inclination of the 
BLR-cloud orbits to the line of sight, and can be calibrated from other independent 
measurements of the same BH systems. BH masses thus estimated are indirect 
and subject to systematics as large as $\sim$0.3~dex \citep[e.g.,][]{gebhardt00b,
greene06,grier13}. Nevertheless, they provide us one of the fundamental parameters 
to study the vast majority of the AGN population. An archetypal LMBH AGN is 
NGC\,4395 \citep{filippenko03}, which is a bulgeless galaxy harboring a central 
BH of \mbh\,$\sim\,3 \times 10^5$~\msun\ measured through reverberation 
mapping observations of the \civ\,$\lambda1549$ line \citep{peterson05}. 
Multi-wavelength observations in the X-ray \citep[e.g.,][]{iwasawa00,moran05,
dewangan08} and Radio bands\citep[e.g.,][]{wrobel06} support the finding of a 
small BH mass in NGC\,4395. Other convincing cases of LMBH in AGNs 
include POX\,52, a dwarf spheroidal galaxy with a BH of $\mbh \sim\,3 
\times 10^5$~\msun\,\citep[e.g.,][]{barth04,thornton08}, and SDSS 
J160531.84$+$174826.1, a dwarf disk galaxy with a BH of $\mbh \sim\,7 
\times 10^4$~\msun\,\citep{dongxb07} .

In general, single-epoch optical spectrum is an effective tool of detecting 
signature of nuclear BHs in active galaxies. AGNs with LMBHs as large sample 
were first searched by \citet{greene04, greene07a} from the Sloan Digital Sky 
Survey \citep[SDSS;][]{york00} yielding $\sim$\,200 broad-line sources with 
the BH masses in the range of $\sim\,10^{5-6}$~\msun, estimated from the 
widths and luminosities of the broad \ha\ emission lines. Independently 
we \citep[][hereafter Dong+12]{dongxb12} carried out a systematic and 
homogeneous search for LMBHs from the SDSS Fourth Data Release 
\citep[DR4;][]{adelman06}, resulting in 309 AGNs, with relatively lower BH 
mass and Eddington ratio distributions. In recent surveys targeting low-mass 
type~2 AGNs, there are more than one hundred sources in nearby dwarf 
galaxies identified as LMBH candidates using narrow-line diagnostics 
\citep[e.g.,][]{barth08,reines13,moran14}. According to \citet{yuanwm14}, if 
the intrinsic Eddington ratio distribution of SMBHs obtained by \citet{schulze10} 
can be applied to the low-mass end, there should exist many more LMBHs 
in the universe. Clearly, a larger LMBH sample than the current ones
homogeneously selected with well-understood completeness is essential to 
construct the intrinsic Eddington ratio and mass functions (Liu et~al. 2018 in 
prep.). These will give more stringent constraints on seed black hole models. 
Moreover, a statistical study of the interplay of LMBH AGNs and their host 
galaxies also requires a larger sample, especially at the lower-\mbh\ and 
lower-accretion-rate end. In this work, we perform an extended search for 
more low-mass AGNs from the SDSS Seventh Data Release \citep[DR7;][]
{abazajian09}, on the basis of our previous work using the SDSS DR4 (Dong+12).

It is difficult to search for AGNs with low-mass black holes since their optical 
spectra are mostly strongly dominated by starlight. Thus careful subtraction of the 
starlight is essential for reliable measurement of the emission lines. In addition, the 
decomposition of the broad and the narrow components of the Balmer lines in LMBH 
AGNs is delicate as the broad lines in the spectra of LMBHs are relatively narrow 
and weak\footnote{ According to Dong+12, the broad \ha\ widths (FWHMs) 
of LMBHs range from 500 to 2200 \kms, with a median of 1000~\kms, which is 
much lower than that of the entire parent broad-line AGN sample ($\approx$3000~\kms)
and even slightly smaller than the traditional demarcation value between AGN 
broad  and narrow lines \citep[1200~\kms; cf.][]{haol05}. }. In order to search 
for LMBHs, we have designed a set of elaborate codes and broad-line selection 
procedures as described in Dong+12. In this work, we follow exactly the same 
method and apply it to the SDSS DR7. Some 204 new LMBHs are found with 
$\mbh < 2 \times 10^6$~\msun\ (to be in accordance with the \citealp{greene07a} 
and Dong+12),  expanding the total SDSS sample of LMBHs to 513. The 
BH masses of the new sample range from $1 \times 10^5$~\msun\ to $2 \times 
10^6$~\msun, with a median of $1.3 \times 10^6$~\msun, and the Eddington 
ratios range from $\sim$\,0.01 to $\sim$\,1. The data analysis and sample selection 
are outlined in Section~2, respectively, and the LMBH sample is described in 
Section~3. In Section~4, the sample properties are discussed, followed by a 
summary in Section~5. Throughout the paper we assume a cosmology with $H_0 = 
70$~\kms~Mpc$^{-1}$, $\Omega_m = 0.3$, and $\Omega_{\Lambda} = 0.7$.


\section{DATA ANALYSIS AND SAMPLE SELECTION}

Our LMBH AGNs are selected following the data analysis procedures described 
in Dong+12, which are only briefly summarized here (see Dong+12 for details). 
We start from the SDSS DR7 spectra classified as ``galaxies'' 
or ``QSOs'' by the SDSS pipeline, excluding objects also in the SDSS DR4. 
The SDSS is a comprehensive imaging and spectroscopic survey using a 
dedicated 2.5\,m telescope \citep{gunn06} located at the Apache Point 
Observatory in Southern New Mexico. It utilizes a wide-field imager \citep 
{gunn98} covering the sky in a drift-scan mode in five filters $ugriz$ \citep 
{fukugita96}, and a 640-fiber-fed pair of multi-objects double spectrographs 
covering the wavelength 3800--9200~\AA\ with a resolution of $\lambda/\Delta 
\lambda \approx 2000$. The diameter of the optical fibers is 3\arcsec\ and the 
instrumental dispersion is $\sim69~\mathrm{km~s^{-1}~pixel^{-1}}$. To ensure 
the \ha\ lying within the wavelength coverage range of the SDSS spectra, only 
sources with redshifts below 0.35 are considered. These result in a parent 
sample consisting of 337,988 ``galaxies'' and 4,697 ``QSOs''. The spectra are 
corrected for the Galactic extinction using the extinction map of \citet{schlegel98} 
and the reddening curve of \citet{fitzpatrick99}, and are then transformed to the 
rest frame with the redshifts provided by the SDSS pipeline.

The spectra of the parent sample are mostly dominated by starlight.  As
the first step, a pseudo-continuum is modeled and subtracted using the
technique described in \citet{zhouhy06}\footnote{Note that there exists
a population of spectra dominated by AGNs, of which the starlight components
are negligible and the broad lines are broad and significant. For the case
we fit simultaneously the AGN power-law continuum together with the 
emission lines including the \feii\ multiplets, the forbidden and the Balmer 
lines \citep[details see][]{dongxb08}.}. The so-called pseudo-continuum is 
a linear combination of several components, including starlight, nuclear 
continuum and the optical \feii\ multiplets. The Balmer continua and high 
order Balmer emission lines are added if it can improve the reduced 
\chisq\ by at least 20\%. The stellar component is 
modeled by six synthesized galaxy spectral templates built up from the library 
of simple stellar populations \citep[SSPs;][]{bruzual03} using the Ensemble 
Learning Independent Component Analysis algorithm \citep{luhl06}, which 
takes all the stellar features into account and can thus significantly mitigate the 
problem of overfitting. A power law is adopted to describe the AGN continuum. 
The optical \feii\ multiplets are modeled by two separate sets of templates 
constructed by \citet{dongxb08,dongxb11} based on the measurements of 
I\,Zw\,1 by \citet{veron04}, one for broad lines and the other for narrow lines.

The next step is to fit the emission lines and to select broad-line candidates.
We focus on the broad \ha\ line (\bha) since it is the strongest broad line in 
the optical spectra of AGNs. Our initial criteria for the addition of a broad 
component of \ha\ are as follows\footnote{Details see Section~3 of Dong+12.}. 
(1) It would result in a significant decrease in \chisq\ with a chance 
probability of F-test $< 0.05$, (2) the width of \bha\  is relatively larger than 
those of narrow lines, particularly \oiii\,$\lambda$5007, (3) \bha\ has a statistically 
significant flux, namely, Flux(\bha) $> 3\sigma$, $\sigma$ is the statistical noise, (4) 
Flux(\bha) $> 10^{-16}$ \flux. After removing the continua, the residual spectra 
are first fitted in order to remove objects without any broad lines according 
to our selection criteria, resulting in a significant reduction of the number of 
objects needing refined fitting ($\sim$303,000\,objects are removed). 
Next, the \ha\ region is initially modeled using pure narrow line profiles without 
broad component. The narrow \ha\ + \nii\,$\lambda\lambda6548,6583$ lines 
are fitted with a narrow-line model built up from the \sii\,$\lambda\lambda6716,6731$ 
doublets, or from the core of \oiii\,$\lambda\lambda4959,5007$ if \sii\ is weak. The 
profiles and redshifts of \ha\ and the \nii\ doublets are assumed to be the same 
as the narrow-line model obtained above. The centroid wavelengths of these 
narrow lines, as well as the flux ratios of the \nii\ doublets $\lambda$6583/$\lambda$6548 
and the \oiii\ doublets $\lambda$5007/$\lambda$4959 are fixed to their theoretical 
values, respectively. Then, a possible additional broad component of \ha\ is 
considered if it satisfies the broad line criteria. This step results in $\sim$17500 
spectra leftover in which a candidate broad \ha\ component may be present.

The broad \ha\ lines of LMBH AGNs are generally relatively narrow and/or weak, 
and their fluxes are susceptible to the subtraction of the narrow lines. Thus for 
the remaining $\sim$17500 spectra, we employ a series of schemes to fit
the narrow \ha\ line, including narrow-line profiles built up from narrow \hb, \sii, 
\oiii\ or \nii, respectively. The result with the minimum reduced $\chi^2$ is adopted 
as the best fit. An example spectrum and the best fitted model are shown in 
Figure~\ref{fig-illustration}.

To ensure the reliability of the existence of a broad component,  we apply further 
a more strict cut on the signal-to-noise ratio of the broad \ha\ flux, \snr(\bha) 
$\geq 5$, where \snr(\bha) is the ratio of the broad \ha\ flux to the total 
uncertainties ($\sigma_{\rm total}$), \snr(\bha) $=$ Flux(\bha) $/\sigma_{\rm total}
$~. $\sigma_{\rm total}$ is the quadrature sum of statistical noise ($
\sigma_{\rm stat}$), the uncertainties arising from subtraction of the continuum 
($\sigma_{\rm cont\_sub}$) and the noise caused by the subtraction of the narrow 
lines ($\sigma_{\rm NL\_sub}$), namely, $\sigma^2_{\rm total} = \sigma^2_{\rm 
stat} + \sigma^2_{\rm NL\_sub} + \sigma^2_{\rm cont\_sub}$~. The $
\sigma_{\rm NL\_sub}$ is estimated using the rest $n$ sets of fitting results that 
are worse than the best one and with the chance-probability of the F-test greater 
than 0.1,
\begin{equation}
\sigma_{\rm NL\_sub} = \sqrt{\frac{ \sum_{i=1}^{n} ({\rm Flux(\bha})_i - 
{\rm Flux(\bha)}_{\rm best})^2}{n}}~.
\end{equation}
\noindent
As for $\sigma_{\rm cont\_sub}$, the term is negligible since the \ha\ absorption 
features are weak in most cases and the continuum decomposition for each 
broad-line AGN candidate has been visually checked to ensure that the subtraction 
error is much below 1~$\sigma$. With the cut of \snr(\bha), we obtain 6092 objects 
with reliable broad \ha\ detections, including 1,850 ``galaxies'' and 4,242 ``QSOs''.


\section{LMBH AGN SAMPLE FROM THE SDSS DR7}

The optical spectra of low-mass AGNs are often substantially contaminated by 
starlight and it is difficult to directly measure the continuum luminosities. Thus the 
luminosities of the broad Balmer lines are adopted to estimate the BH masses. 
In this study, for ease of comparison, the BH masses are calculated using the 
formalism presented in GH07 following Dong+12,
\begin{equation}
\mbh = (3.0^{+0.6}_{-0.5}) \times 10^6
\left(\frac{\lbha}{10^{42}~{\rm erg~s^{-1}}} \right)^{0.45 \pm 0.03}\left(
\frac{\fwbha}{10^{3}~\mathrm{\kms}} \right)^{2.06 \pm 0.06} \msun.
\end{equation}
\noindent
The formalism was derived using the empirical correlations of \lfive--\lbha\ and 
$\rm FWHM_{\bha}$--$\rm FWHM_{\bhb}$ from \citet{greene05b}\footnote{In 
fact, \citet{greene05b} gives the correlation between \lfive\ and the combined 
\ha\ luminosity of both broad and narrow components. However, as described 
in \citet{greene05b}, ``the best-fit parameters are virtually unchanged when 
only the broad component is considered''. In addition, the narrow \ha\ component 
may be contaminated by the host galaxies for our LMBHs. Thus in this study 
\lfive\ is estimated using merely broad \ha\ component.}, along with the 
radius-luminosity relation reported by \citet{bentz06}. Finally a sample of 204 
objects with BH masses less than $2 \times 10^6$~\msun\footnote{We adopt 
the same upper limit of LMBHs as in GH07 and Dong+12 for consistency.} are 
obtained. Table~\ref{tab-basic} summarizes the basic data of the sample. The 
emission line parameters obtained from the best-fit models as described in 
Section~2.1, are tabulated in Table~\ref{tab-emline} and Table~\ref{tab-mbh}, 
along with the BH mass and luminosity. The flux of \feii\,$\lambda4570$ is 
calculated by integrating the flux density of the corresponding \feii\ multiplets 
from 4434~\AA\ to 4684~\AA\ in the rest-frame. 

Our sample has a median redshift of 0.1 (see Figure~\ref{fig-redshift}) and BH 
masses ranging from $1.1 \times 10^5$~\msun\ to $2.0 \times 10^6$~\msun. 
The FWHMs of \bha\ span a range of $\sim$500--2200 \kms, with a median
of 1000~\kms. The broad Ha line luminosities \lbha\ are in the range of  $\sim
10^{39.3}$--$10^{42.1}$\lum with a median lying at $10^{39.3}$\lum. 
We caution that dust extinction may attenuate the observed broad \ha\ flux 
for some of the low-mass AGN and may lower somewhat the measured \lbha.  
Given that the measurement of the broad \hb\ line in low-mass AGN suffer 
large uncertainties, it is difficult to quantify the dust reddening effect. Following 
GH07 and Dong+12, we do not correct for the dust reddening in this work.
The Eddington ratio (\llambda) is defined as the ratio of the bolometric 
luminosity ($L_{\rm bol}$) to the Eddington luminosity ($\ledd  = 1.26 \times 
10^{38} \mbh/\msun$). $L_{\rm bol}$ is derived from the optical luminosities 
at 5100~\AA\ using a bolometric correction factor of 9.8 \citep {mclure04}, 
$L_{\rm bol}=9.8$\lfive, where \lfive\ is derived from the broad \ha\ luminosity 
using the scaling relation given in \citet{greene05b}. The Eddington ratios 
thus estimated for the current LMBH sample range from 0.01 to 2. Figure~\ref 
{fig-ourdist} and Figure~\ref{fig-optpara} show the distributions of BH masses 
and Eddington ratios, as well as the luminosities and FWHMs of broad \ha, for 
our sample. Those of the Dong+12 and GH07 samples are also plotted for 
comparison.

As expected, the distributions of the DR7 LMBHs are similar to those in 
Dong+12. The median \mbh\ of the present sample is $1.3 \times 10^6$~\msun, 
comparable to $1.2 \times 10^6$~\msun\ of Dong+12. As for \llambda, the 
medians are $-$0.59 and $-$0.64 in log-scale, respectively. The medians 
of \lbha\  and \fwbha\ for the current sample and Dong+12 are very close 
(41.00 versus 40.99 and 2.99 versus 3.02, calculated in logarithm, respectively). 
In addition, the standard deviations of these quantities, including \mbh, \llambda, 
\lbha\ and \fwbha, of the two samples are consistent within 0.07~dex. 


\section{SAMPLE PROPERTIES}
In combination with the 309 objects in Dong+12, we expand the SDSS LMBH 
AGN sample to a total of 513 sources up to the DR7. This is the largest sample 
of low-mass AGNs so far, which have the uniform and homogeneous selection 
criteria and well measured AGN parameters, thanks to the homogeneity and 
accurate spectrophotometry of the SDSS. It is worthwhile to investigate the 
statistical properties for the total sample. In this section we present some of the 
ensemble properties of the LMBH AGNs and their host galaxies based on the 
SDSS data, as well as data from X-ray and Radio surveys.

\subsection{Narrow Line Diagnostic Diagrams}

Compared with \hii\ galaxies, AGN can emit a harder continuum which results 
in a distinct ionization condition in surrounding gas. Specific emission line 
ratios can help distinguish the central radiating sources that ionize the 
circumnuclear medium. In practice, two-dimensional diagrams of certain 
narrow-line ratios have been widely applied to discriminate between \hii\ 
galaxies and type 2 AGNs \citep[e.g.,][]{baldwin81, veilleux87, ho97a, kewley01, 
kewley06, kauffmann03b}. The so-called BPT diagrams involving the narrow 
line ratios of \ha, \hb, \oiii, \nii, \sii, and \oi\ are shown in Figure~\ref{fig-bpt}. 
In general, the distributions of our LMBHs are consistent with those of Dong+12. 
On the \oiii\,$\lambda5007/$\hb\ versus \nii\,$\lambda6583/$\ha\ diagram 
(panel (\textit{a})), the vast majority of these LMBHs are located in the region 
of either Seyfert galaxies or composite objects, and only a few sources 
($\sim$\,10) fall into the pure star-forming region. On the diagrams of the 
\oiii~$\lambda5007/$\hb\ versus \sii~$\lambda\lambda6716,6731/$\ha\ 
and \oi~$\lambda6300/$\ha\ diagrams (panels (\textit{b}) and (\textit{c})), 
about two-thirds of these objects are found in the AGN region, while the rest 
are located in the \hii\ region.

These results confirm the AGN nature for the vast majority of our sample, 
though a few objects have the narrow-line spectra similar to those of star-forming 
galaxies. Regarding those objects located in the pure \hii\ portion, their broad 
\ha\ fluxes, FWHMs and luminosities are all similar to those of the whole parent 
sample, with chance probabilities $>0.1$ according to Kolmogorov--Smirnov test 
(KS test). Visual checks of their SDSS spectra and the statistics (broad \ha\ 
flux with $\rm S/N > 5$) both demonstrate that the broad \ha\ components of these 
objects are significant and reliable. A probable explanation is that the spectra 
of these objects are strongly contaminated by star light from those star-formation 
regions in the host galaxies, given the relatively large aperture of the fiber of 
3\arcsec. 

We note that about two dozens of objects fall into the region of low-ionization 
nuclear emission-line region sources \citep[LINERs;][]{heckman80} according 
to the Seyfert--LINER demarcation lines of either \oiii~$\lambda5007/$\hb\ $= 
3$ in panel (\textit{a}) or \citet{kewley06} in  panels (\textit{b}) and (\textit{c}). 
LINERs are commonly found in early-type galaxies with classical bulges 
containing supermassive BHs especially with old stellar populations \citep 
{ho97c}. LMBHs tend to reside in late type disk-dominated galaxies often 
harboring pseudobulges just on the opposite. Therefore such LMBH LINERs 
are of great interest to study LINERs as a population though it is not a surprise 
that only a small number of low-mass AGNs show LINER-like spectra. Mostly 
manifesting themselves as low-luminosity AGNs, LINERs are thought to have 
relatively low accretion rates and their accretion flows are dominated by ADAF 
\citep[Advection-Dominated Accretion Flow; see e.g.,][]{kewley06,ho09,narayan95}. 
The combination of low BH mass and low accretion rate make it even difficult 
to identify LINERs with LMBHs in the presence of host galaxy light. These may 
explain the low incidence of LINERs in the LMBH sample. 

\subsection{X-ray Properties}

X-ray is an important bandpass to study BH accretion. A number of LMBH 
AGNs in the GH07 sample have been observed in X-ray with \chandra\ 
\citep[e.g.,][]{greene07c, desroches09, dongrb12}, which gave a snapshot 
for the X-ray properties of LMBHs. Focusing on low-mass AGNs with very low 
Eddington ratios (\lratio\,$\sim 0.01$), i.e., the faintest AGN population known, 
\citet{yuanwm14} studied the X-ray properties of four objects observed by 
\chandra\ in the Dong+12 sample, and suggested there should exist a large 
population of underluminous LMBHs in the nearby universe. \citet{plotkin16} 
performed a similar analysis on seven low-mass black holes with low Eddington 
ratios from GH07 using \chandra\ observations. The two well-studied LMBHs, 
NGC~4395 and POX~52, were found to show rapid and strong X-ray 
variabilities \citep[e.g.,][]{iwasawa00,moran05,dewangan08,thornton08}. Their 
spectral characteristics resemble those classical Seyfert~1 galaxies. \citet 
{miniutti09} performed a detailed analysis on four LMBHs from \citet{greene04} 
using \newton\ observations, finding that they are extremely variable and three 
out of the four objects show soft excess in their spectra. Similarly, \citet 
{ludlam15} followed a study with 14 LMBHs from GH07 with \newton, eight of 
which show soft excess emissions. In addition, LMBHs also present diverse
timing properties. In general, LMBHs tend to show larger X-ray variability 
amplitude than their more massive cousins. \citet{panhw15} and \citet{ludlam15} 
showed the previous known inverse correlation between \mbh\ and the normalized 
excess variance of X-ray variability, and flattens at \mbh\,$\sim10^6$~\msun\ 
and thus vanishes for LMBH AGNs.

In this study, we briefly investigate the statistical X-ray properties of the total
sample of 513 sources based on the \rosat\ All-Sky Survey \citep[RASS,][]
{boller16} and pointed observations (2RXP). Detailed X-ray spectral and timing 
analysis using \newton\ are deferred to later work.

\subsubsection{X-ray Detection}
Of the 513 sources, 85 were detected by the RASS and 32 in pointed 
observations, with 15 detected in both (for these sources, the data from the 
pointed observations are adopted). Thus a total of 102 objects have X-ray 
detections with the \rosat\ Positional Sensetive Proportional Counter (PSPC).
Their X-ray fluxes and spectral indices in the 0.1--2.4 keV band are estimated
following \citet{schartel96} and \citet{yuanwm98,yuanwm08}, assuming an 
absorbed power-law spectral shape. As the first step, the photon index $
\Gamma$ is calculated from the two hardness ratios\footnote{The two
hardness ratios are defined as HR1 = (B - A)/(B + A), HR2 = (D - C)/(D + C), 
where A, B, C and D are the number of source counts in the energy channels
of 11--41, 52-201, 52--90 and 100--201 respectively.} if available. The 
absorption column density \nh\ is fixed at the Galactic value \citep{dickey90}, 
or set to a free parameter if no meaningful $\Gamma$ is obtained using the 
Galactic \nh. For those sources without meaningful hardness ratios, the 
mean value ($\Gamma = 2.36$) is adopted. Next we estimate the X-ray fluxes 
in the 0.1--2.4 keV band from the count rates using the energy-to-counts 
conversion factor (ECF; \rosat\ AO-2 technical appendix, 1991), which is 
calculated from the \rosat\ PSPC effective area for each source, for the given 
power-law spectrum with $\Gamma$ and \nh\ obtained above. The X-ray 
fluxes in the band of 0.5--2.0 keV are tabulated in Table~\ref{tab-rosat}, 
which range from $1.8 \times 10^{-14}$ to $2.9 \times 10^{-12}$ \flux. The 
corresponding luminosities span a range from $8.0 \times 10^{39}$ to $1.1 
\times 10^{44}$ \lum, comparable to $9.3 \times 10^{39}$--$6.9 \times 
10^{43}$ \lum\ for GH07 sample. This indicates that the X-ray emission is 
mostly dominated by AGN radiation, since the X-ray luminosities of normal 
galaxies are generally below this level.

For those undetected by the \rosat, upper limits on the X-ray fluxes are 
estimated using the method in \citet{yuanwm08}. An upper limit of 12 source
counts is adopted for each undetected object, and thus the count rate upper
limit is calculated using the corresponding effective exposure time from the 
RASS exposure map. Then the flux upper limits are estimated using the method 
described above assuming the Galactic \nh\ and the mean $\Gamma$ of 2.36.

 
\subsubsection{X-ray versus \oiii\,$\lambda$5007 Luminosity}

The soft X-ray emission is susceptible to absoprtion, while the \oiii\ luminosity 
is suggested to be an isotropic indicator of the intrinsic AGN power since the 
\oiii\ line originates from the narrow-line region and should be unaffected by 
obscuration. A strong correlation between the X-ray and \oiii\,$\lambda$5007 
luminosity has been found in unobscured AGNs \citep[e.g.][]{panessa06}. 
This can help to discriminate whether our low-mass AGNs are heavily 
obscured.

The relation between the 2--10 keV luminosity and \oiii\,$\lambda$5007 
luminosity for the X-ray detected sources is shown in Figure~\ref{fig-optLx}, 
overplotted are the low-mass AGN sample with \rosat\ detections from GH07, 
LMBHs observed by \chandra\ from \citet{dongrb12}, more massive AGNs 
in \citet{jincc12} and low-mass AGNs with log $(L_{\rm bol}/L_{\rm Edd}) < -1.5$ 
observed using \chandra\ in \citet{yuanwm14} and \citet{plotkin16}.
Luminosities in the 2--10 keV band are calculated by extrapolating the spectra 
of the \rosat\ to 10~keV assuming a photon index of 2.5. The 
\rosat\--detected sources in our total sample are roughly consistent with the 
correlation between X-ray and \oiii\,$\lambda$5007 luminosity derived by more 
massive AGNs \citep{panessa06}. It indicates that our low-mass AGNs with 
X-ray detections suffer little or no obscuration in X-rays. 


\subsubsection{Optical--X-ray Spectral Index}

The optical/X-ray effective spectral index, \aox$\equiv -0.384\,$log($f_{\rm 
2500 \angstrom}/f_{\rm 2 keV})$, where $f_{\rm 2500 \angstrom}$ and 
$f_{\rm 2 keV}$ are the rest-frame flux densities at 2500~\AA\ and 2~keV 
respectively, is commonly used to describe the relative dominance of the 
optical and X-ray emission \citep{tananbaum79}\footnote{The definition is 
different from the original one in \citet{tananbaum79} by a negative sign.}. 
The statistical properties of \aox\ have been well studied for the 
classical AGNs \citep[e.g.][]{avni86,yuanwm98,vignali03,steffen06}. 
Recently, \citet{dongrb12} studied the \aox\ properties of 49 low-mass AGNs 
in the GH07 sample with \chandra\ observations. In this work, we briefly 
investigate the \aox\ properties of LMBHs based on \rosat\ observations.

The \aox\ indices are calculated as follows. $f_{\rm 2 keV}$ is derived 
from the \rosat\ observations as described in Section~4.2.1 and $f_{\rm 2500 
\angstrom}$ is calculated from \lbha\ using the relation between \lbha\ and 
\lfive\ given by \citet{greene05b} assuming a spectral index of 1.56 \citep[$f_{\lambda} 
\propto \lambda^{-1.56}$;][]{vandenberk01}. The \aox\ values are listed in 
Table~\ref{tab-rosat}, and their distribution is shown in Figure~\ref{fig-aoxdist}. 
The \aox\ distribution has a large scatter, ranging from $-$1.58 to $-$0.70, 
with a median of $\langle$\aox$\rangle = -1.12$, which is systematically flatter 
than those of AGNs with more massive black holes \citep[e.g., \aox\,$\sim-1.5$;]
[]{yuanwm98}.  The \aox\ distribution of our low-mass is similar with that of the 
\rosat\--detected sources in GH07, but slightly flatter than that of the LMBHs 
with ${\emph Chandra}$ observations in \citet{dongrb12}. These LMBHs are 
roughly in accordance with the relation between \aox\ and the monochromatic 
luminosity at 2500~\AA\ defined by those more massive AGNs \citep {steffen06}, 
albeit with large scatter (see Figure~\ref{fig-aoxl2500}). In addition, Figure~\ref{fig-aoxl2500} 
shows that low-mass AGNs have systematically flatter \aox\ and a larger scatter
compared to those of more massive black holes, which is consistent with that 
in \citet{dongrb12}.

As can be seen from Figure~\ref{fig-aoxl2500}, though the distribution of our 
LMBH AGNs are mostly confined within the 1~$\sigma$ scatter of the \aox--$L_{\rm 
2500\angstrom}$ relationship, there is a much larger scatter when the \chandra\ 
samples of \citet{dongrb12} and \citet{yuanwm14} are taken into account, which 
contains some X-ray weak objects detected by \chandra\ owing to its high sensitivity. 
To explore further the large scatter of \aox\ for LMBH AGNs, we plot in 
Figure~\ref{fig-aoxmbh} \aox\ versus the Eddington ratio and BH mass for our sample 
as well as the other AGN samples as in Figure~\ref{fig-aoxl2500}. No significant 
correlation is found between \aox\ and  \llambda\ over a wide range 
for the LMBHs spanning nearly two orders of magnitudes. This is consistent with 
previous results for LMBH AGNs \citep[e.g.,][]{greene07c,dongrb12}. Regarding 
the \aox--\mbh\ relationship, no correlation is found for LMBHs only given the 
narrow dynamic range of \mbh; however, a weak correlation (Spearman's rank 
correlation coefficient is $-$0.25 with a null probability of $\sim\,10^{-7}$) appears 
to be present when taking into account AGNs with more massive BHs from the 
other samples, albeit the large dispersion for the LMBHs. A similar weak correlation 
was also suggested by \citet{dongrb12}. Thus the wide \aox\ distribution for LMBH 
AGNs is not caused by the distributions of the Eddington ratio or BH mass.

Three possibilities might account for the large scatter in the \aox\ distribution.
The first is X-ray variability. It has been demonstrated that the amplitude of short 
time-scale X-ray variability is anti-correlated with the BH mass for AGNs with $\mbh 
> 10^6$~\msun\ \citep[e.g.,][]{ponti12,kelly13}, below which the relation flattens 
\citep{panhw15,ludlam15}. This means that LMBH AGNs show the largest amplitude 
of X-ray variability among all AGNs. Moreover, LMBH AGNs appear to show strong
variations in the X-ray spectral shape, as manifested by their wide range of the 
spectral indices ($\Gamma$ = 1.2--4.0).  A typical example is NGC 4395, which was
found to exhibit rapid and strong X-ray variability \citep{iwasawa00,dewangan08}, 
and its X-ray spectral slope varied from $\Gamma < 1.25$ to $\Gamma > 1.7$ in 
about one year \citep{moran05}. Similar behavior was also found in POX 52
\citep{thornton08}.

Secondly, LMBHs may have a wide distribution in the intrinsic X-ray luminosities. 
As commonly believed, the X-ray emission is mainly produced in the hot corona
by inverse-Compton scattering of optical/UV photons from the disk. The relative 
dominance of the optical/UV and X-ray emission is determined by the fraction of 
the energy deposited into the hot corona out of the total viscos energy produced 
in the accretion disk. For instance, in models that the corona is heated by magnetic 
re-connection \citep{liubf03} , the fraction is mainly determined by the magnetic 
field strength in the corona, which may vary from one object to another. 

Thirdly, X-ray absorption may play a role as well. Although no strong effect of X-ray 
obscuration is present for most of the objects as suggested by the 
$L_{\rm 2-10keV}$--$L_{\rm \oiii}$ relation above, moderate absorption cannot be ruled out 
in some of the objects. It is probable that the environment of LMBH may be more 
gas-rich compared to normal AGNs powered by more massive BHs, whose stronger 
radiation makes it easier to expel circumnuclear materials.

\subsection{Radio Properties}

We explore the radio properties for our total LMBH sample using radio data at 
20~cm from the VLA FIRST survey\footnote{The VLA FIRST aims to produce 
Faint Image of the Radio Sky at Twenty centimeters using Very Large Array, 
which is operated by the National Radio Astronomy Observatory.} \citep{becker95}. 
There are 26 objects detected, leading to a low detection fraction of 5\%. The 
low incidence of radio activity is consistent with the result in GH07. The radio 
powers at 20~cm of these sources range from $2 \times 10^{21}$ to $4 \times 
10^{23}$ \whz, with a median of $4 \times 10^{22}$ \whz (see Table~\ref{tab-first}). 
Figure~\ref{fig-radiop6cm} shows their radio power versus the \oiii\ luminosity, 
which appears broadly consistent with what is found in GH07.

As a common practice, the radio loudness parameter, defined as the radio to 
optical flux ratio (\rkel\ $\equiv f_{\rm 6 cm}/f_{\rm 4400 \angstrom}$), is used 
to separate radio-loud AGNs from radio-quiet ones, with an operational dividing 
value of 10. The radio loudness is calculated for the radio-detected sources as 
following: $f_{\rm 6 cm}$ is derived assuming a radio spectral index of $\alpha_{r}$ 
= 0.46 \citep[$f_{\nu} \propto \nu^{-\alpha_{r}}$;][]{ulvestad01} from the flux 
density at 20~cm; while $f_{\rm 4400 \angstrom}$ is estimated in the same way 
as $f_{\rm 2500 \angstrom}$. Among the 26 sources with FIRST detections, 
23 are radio loud, corresponding to a radio loud fraction of 4\%. For those 
sources undetected in FIRST\footnote{Note that $\sim$30 sources are not covered 
by the FIRST, thus their upper limits are not calculated.}, upper limits on radio 
loudness are estimated assuming a radio flux density of 1~mJy at 20~cm, which 
is the detection limit of the FIRST survey(see Figure~\ref{fig-radioloudness}).

The radio loud fraction in the LMBH AGN sample is lower than that of classical
AGNs \citep[$\sim$15\%;][]{ivezic02}. This seems to be consistent with the 
finding that radio sources may prefer to reside in more massive galaxies \citep 
{best05}. For our LMBHs, their host galaxies are indeed at the lower-stellar-mass 
end of the parent sample from the SDSS DR7 (see Section~5.3 for details). 
However, it should be noted that, the above radio loud fraction should only be 
considered as a lower limit, since the FIRST flux sensitivity is not deep enough 
to give stringent constraints on the radio loudness. Further deep and high 
spatial resolution radio surveys are needed to determine the true radio loud 
fraction for LMBHs.

As a natural interpretation, the radio emission of an active galaxy is a mixture of 
contribution from nuclear AGNs and star formation in the host galaxy. This may 
be particularly true for the fluxes list in the FIRST catalog we used here, which 
are obtained by fitting the sources usually as extended radio sources (cf. the 
`Deconv.MajAx' column in the catalog). While it is difficult to seek clues 
from the radio morphology since most of these sources have very weak radio 
emissions ($S_{20cm}< \sim$\,5~mJy, close to the detection limit of 1~mJy), 
and their FIRST images are unresolved. Only one brighter object ($S_{20cm}
>$\,10~mJy), SDSS J122412.51$+$160012.1, probably shows marginally 
extended structure with  integrated to peak flux ratio of 1.5 and deconvolved 
major axis = 5.45\arcsec. On the other hand,  radio luminosities at 6~GHz 
may provide some information on the radio origins. It is suggested that for AGNs 
at low redshifts, low-luminosity radio emission with $L_{\rm 6 GHz} < 10^{23}$ 
\whz\ seems to be powered by star formation, while that with $L_{\rm 6 GHz} 
\geq\ 10^{23}$ \whz\ is more likely AGN-dominated \citep[e.g.,][]{kimball11,
condon13,kellermann16}. The luminosity at 6~GHz is derived assuming a 
spectral slope of 0.46. All the objects except one (SDSS J122412.51$+$160012.1)
have 6~GHz luminosities located in the range of $10^{21}$ \whz\ $\leq L_{\rm 
6 GHz} \leq 10^{23}$ \whz, and thus their radio fluxes may also be contributed 
from host stellar processes.

\subsection{Comparison with NLS1s}
Most low-mass AGNs tend to have broad line widths (FWHMs) narrower than 
$\sim$2000~km/s, which is the conventional criterion of narrow-line Seyfert~1 
galaxies \citep[NLS1s;][]{osterbrock85}. It is thus interesting to compare
these two AGN sub-classes. In fact, some objects in our total sample have also 
been classified as NLS1s in a comprehensive study of NLS1 AGNs from the 
SDSS DR3 by \citet{zhouhy06}. NLS1s are characterized by some peculiar 
properties including the strong \feii\ lines, weak \oiii$/$\hb\ line ratios, high 
Eddington ratios, low radio loud fraction, significant soft X-ray excess (in some), 
strong X-ray variability and steep X-ray spectra \citep[see, e.g.,][for reviews] 
{laor00, komossa08}. Based simply on the BH mass estimation method from 
the line width and luminosity, a relative narrow broad line width can result from 
either a low black hole mass, or/and a high Eddington ratio. With their peculiar 
multi-wavelength properties, NLS1s are found to cluster at one end of the 
so-called EV1 which are suggested to be driven by high accretion rates 
\citep{boroson92}. Thus NLS1s are generally suggested to harbor relatively
less massive black holes radiating near their Eddington limits.

However, LMBHs are selected only from black hole masses regardless the 
Eddington ratios. They are simply the low-BH-mass counterparts of classical 
Seyfert galaxies, and are expected to exhibit diverse properties depending on 
the Eddington ratios, which span a wide dynamical range. \citet{greene04} 
found that their low-mass AGNs span a larger range in both the \feii\ and the 
\oiii\ strengths relative to \hb\ than classical NLS1s. There is a wide distribution 
of the Eddington ratios in our sample, some as low as two orders of magnitudes
below their Eddington limits. LMBHs show diverse properties in X-ray, with a 
wide range of the X-ray photon indices (e.g. $\Gamma$ = 1.0--2.7 in \citealp
{desroches09}; $\Gamma$ = 1.5--2.8 in \citealp{dongrb12} and $\Gamma$ 
= 1.2--4.0 in the soft X-ray band for our sample). Moreover, some LMBHs, 
especially those with low Eddington ratios, do not show soft X-ray excess, as 
found in the spectrum of NGC 4395. On the other hand, some typical NLS1s, 
as those with extremely narrow widths of the broad lines as studied by 
\citet{aiyl11}, do have BH masses as low as $10^6$~\mbh\ or below.  

It is clear that NLS1s, when solely selected from the line width, are a heterogeneous 
class of AGNs which may include ordinary Seyfert galaxies having simply LMBHs, 
regardless the Eddington ratios. Classical NLS1s were selected by also considering 
the strong \feii\ and weak \oiii\ emissions. These tend to select objects with high 
Eddington ratios given the observed Eigenvector \Rmnum{1}\ correlations. We thus 
suggest that a simple criterion based solely on the line width may not be a good 
approach for the selection of NLS1s and other criteria have to be incorporated. 
Similar remarks can also be found in \citet{aiyl11}.

One interesting feature is that both types have relatively low fraction of radio-loud
objects. This might be related to the suggested observed statistical relations 
among radio loudness, BH mass and Eddington ratio. However, the underlying 
physical mechanisms driving these relations are not known.

\section{HOST GALAXY PROPERTIES}
In this section, we briefly discuss on the properties of the host galaxies for the 
LMBHs and the possible co-evolution with BHs by making use of the SDSS data.  

\subsection{Luminosities, Colors and Morphologies}
The host galaxy luminosities are calculated in a way following Dong+12 and GH07. 
First, the AGN luminosities are estimated from the broad \ha\ luminosities\footnote{
Assuming an optical spectral shape of 1.56 \citep[$f_{\lambda} \propto \lambda^{-1.56}$;][]
{vandenberk01}, we obtain the AGN luminosities using the scaling relation between 
\lbha\ and \lfive\ described in \citet{greene05b}.} that are free from starlight contamination, 
and are then subtracted from the SDSS photometric Petrosian $g$-band magnitude. 
The host galaxy luminosities are then derived after $K$-corrections using the routine 
of \citet{blanton07}. The distributions of the $g$-band absolute magnitudes of AGN, 
host and the total are shown in Figure~\ref{fig-mag}, with those in Dong+12 and GH07 
plotted for comparison. The distributions of these luminosities for the DR7 LMBH sample 
are similar to those in Dong+12. The current sample has a median AGN luminosity 
of $M_g = -17.80$~mag and a median host galaxy luminosity of $M_g = -20.14$ mag 
(these values are $-$17.70 and $-$20.22 respectively in Dong+12). The median host 
galaxy luminosity is slightly brighter than the characteristic luminosity of $M^{*}_{g} = 
-20.1$~mag at $z=0.1$ \citep{blanton03}. Only 6 objects have host galaxy luminosities 
falling into the dwarf galaxy of $M_g > -$18.0~mag. It indicates that although LMBHs 
usually indeed reside in small stellar systems, their hosts may not necessarily be
dominatly dwarf galaxies. Note that the host galaxy luminosities are possibly overestimated 
in our AGN-host separation method since the fiber loss may result in underestimation of 
the nuclear luminosity, though the systematic overestimation is smaller than 0.3~mag 
according to GH07. Furthermore, the AGN luminosities estimated from the \ha\ luminosities 
assuming a power-law continuum may be subject to an uncertainty of about 0.1~mag 
(Dong+12).

In general, it is not practical to visually classify the galaxy morphology for most 
of our sample objects using the SDSS images due to their limited depth and spatial 
resolution. Instead we try to gain information on the morphologies from the host 
galaxy colors. The $u-g$ colors are found to have a mean value of 1.28 mag 
with a standard deviation of 0.38 mag, and the $g-r$ colors have a mean of 
0.61~mag and a standard deviation of 0.16~mag. They correspond to the typical 
colors of Sbc galaxies according to \citet{fukugita95}.

\subsection{Stellar Populations}
Two stellar indices, the 4000~\AA\ break strength ($D_{4000}$) and Balmer 
absorption line index (H$\delta_{\rm A}$) are used to infer the stellar populations 
of the host galaxies. In general, $D_{4000}$ is small for young stellar populations 
($D_{4000} < 1.5$ for ages $<1$~Gyr ) and large for old, metal-rich galaxies. 
On the other hand, galaxies having experienced star burst activity that has 
ended $\sim$0.1--1 Gyr ago tend to show strong H$\delta$ absorption lines 
\citep{kauffmann03a}. Hence these two indices can be used to constrain the 
mean stellar ages of the host galaxies of LMBHs and to diagnose the star 
formation history over the past few Gyrs.

The two indices are calculated from the decomposed stellar component as 
described in Section~2.1 based on the definition and calculation method in \citet 
{kauffmann03a}. Note that only 157 objects with the AGN contribution less 
than 75\% at 4000~\AA\ in the SDSS spectra are used. The resulting typical 
errors (1~$\sigma$) of the two indices for the LMBH sample are 0.1 and 
1.6~$\AA$, respectively (see Section 4.2.3 in Dong+12 for the details of error 
estimation).

The distribution of 419 low-mass AGNs (262 objects in Dong+12 also included) 
on the $D_{4000}$--H$\delta_{\rm A}$ plane is shown in Figure~\ref{fig-d4k}, and 
the distribution of $\sim$492,000 inactive galaxies is plotted as contours for 
comparison. These inactive galaxies are derived from the SDSS DR7 classified 
as ``galaxies'' by the SDSS pipeline and with $\rm S/N$ $\rm > 5\,pixel^{-1}$ around 
4000~\AA\ in their spectra. In addition, broad-line AGNs and those Seyfert~2
galaxies defined by the BPT diagram are excluded. Note that the negative values 
of H$\delta_{\rm A}$ are due to the definition and corresponding calculation method 
of the line index \citep{kauffmann03a, worthey97}. Most of our LMBH hosts 
(112 out of 157) have $D_{4000} < 1.5$ indicating that their mean stellar ages tend 
to be less than 1~Gyr. In general, for those galaxies with $D_{4000} < 1.5$, if they 
have experienced instantaneous burst of star formation in the last few Gyrs, they 
tend to have H$\delta_{\rm A} > 5~\AA$. However, most objects with $D_{4000} 
< 1.5$ in our LMBH sample do not follow this tendency. It indicates that the distribution 
of our LMBHs on the $D_{4000}$--H$\delta_{\rm A}$ plane is more likely overlapping 
the locus of continuous star formation, which is in broad agreement with that of normal 
galaxies. 

Studies of optical images and spectra indicated that there are two distinct 
populations of the host galaxies of the LMBHs \citep{greene08,jiangyf11}. The 
major are gas-rich, late-type galaxies similar to the prototypal NGC~4395. The 
rest are spheroidal galaxies like POX~52 which has deficient gas and red color. 
\citet{kormendy12} suggested that such spheroidal galaxies were transformed from 
irregular and late-type disk galaxies at an early epoch, by both internal and secular 
processes. The low fraction of LMBH host galaxies with nearby companions also 
imply these galaxies mainly evolve through secular processes \citep{jiangyf11}. 
This is also supported by the above finding that most of the host galaxies of our 
LMBHs undergo continuous star formation in the past few Gyrs.

\subsection{Stellar Masses}
The stellar masses of the host galaxies are estimated using three methods. The 
first is by using Near-infrared (NIR) photometry which can be considered as a 
tracer of stellar mass. In the NIR band, the luminosities of the LMBH AGNs are 
dominated by host galaxies, and AGNs have negligible contribution 
($\sim7$\%\footnote{We simply estimate the AGN contribution in the 
K-band using the relative flux ratio between 5100~\AA\ and 2.2~micron from 
the mean nuclear spectral energy distribution (SED) of the nearest Seyfert~1 
galaxies \citep{prieto10} that includes the AGN continuum and torus emission 
in the NIR band. The flux at 5100 \AA\ is derived from \lbha\ \citep{greene05b}.}).  
Moreover, the mass-to-light ratio in the NIR band is sensitive neither to dust 
absorption nor to star formation history. Of the total sample, 180 objects are 
detected in the Two-Micron All-Sky Survey (2MASS\footnote{Our LMBHs 
tend to be extended sources since they are mostly located in the low-redshift 
universe. Thus the result is derived by cross-matching the LMBH sample with 
2MASS All-Sky Extended Source Catalog (XSC).}) and among the rest, 
129 objects are detected in the UKIRT Infrared Deep Sky Surveys \citep[UKIDSS;]
[]{lawrence07}. Here we adopt the mass-to-light ratios provided by \citet[the scatter 
is about 0.1~dex]{into13}, which are dependent on the galaxy colors $r - i$,
\begin{equation}
{\rm log}(M_*/L_{Ks}) = 2.843(r-i) - 1.093,
\end{equation}
where $L_{Ks}$ is the Ks-band luminosity from 2MASS in units of $L_{\odot}$, 
and $r - i$ is the color derived from the SDSS Petrosian magnitudes after 
$K$-correction and subtraction of the AGN continuum. The $K$-correction is 
calculated following \citet{blanton07}. The K-band magnitudes of the 
UKIDSS-detected sources are transformed to the 2MASS Ks-band using the 
color equations given by Hewett et al. (2006). For the sources not detected 
in 2MASS and UKIDSS, we derive their stellar masses from the MPA--JHU
catalog\footnote{The MPA--JHU catalog, which contains properties of millions 
of galaxies from the SDSS DR7, was produced by a collaboration of researchers 
from the Max Planck Institute for Astrophysics (MPA) and the Johns Hopkins 
University (JHU). It contains two subsets, the raw data and the derived data. 
The raw data contain spectral parameters including the line fluxes, equivalent 
widths and continuum indices, as well as basic information about the objects 
such as their redshifts, velocity dispersions and plate, fiber and MJD. The 
derived data include gas-phase metallicities, star formation rates and stellar 
masses. The stellar masses are calculated from fits to the photometry with 
population synthesis models following the philosophy of \citet{kauffmann03b} 
and \citet{salim07}. In this study we use the improved version of stellar mass 
catalog which can be obtained from \url{http://home.strw.leidenuniv.nl/~jarle/SDSS/}.} 
\citep {kauffmann03b, brinchmann04, tremonti04, salim07} if available, or 
otherwise using the relation between $M_*$ and the 3.4~$\mu$m luminosity 
from the Wide-field Infrared Survey Explorer (WISE) All-Sky survey calibrated 
using the MPA--JHU catalog \citep{wenxq13}. We compare the stellar masses 
given by the MPA--JHU catalog with those derived from the NIR luminosities 
and $r-i$ colors for those objects detected in both surveys, and found 
that the two estimators are statistically in good agreement. The stellar masses for 
the total sample range from $6.0 \times 10^8$~\msun\,to $2.5 \times 
10^{12}$~\msun, with a median of $3.7 \times 10^{10}$~\msun, slightly lower 
than that of the MPA--JHU sample ($5.7 \times 10^{10}$~\msun, see 
Figure~\ref{fig-stellarmassdist} for their distributions). The stellar masses are 
mostly greater than $10^{9.5}$\msun, which is the mass of the Large Magellanic 
Cloud (LMC), indicating again that LMBHs do not necessarily reside in dwarf galaxies.

The 512 sources with reliable stellar mass estimations are plotted on the 
color--mass diagram (see Figure~\ref{fig-stellarmasscolor}). The green lines 
indicate the location of the so-called green valley defined by \citet{schawinski14}, 
which is a superposition of red and blue galaxies with the same intermediate optical 
colors. It has been suggested to be a special stage in the evolution of massive 
galaxies after star formation is quenched, and perhaps as an evidence of AGN 
feedback. Our LMBH sources span almost the entire $u-r$ color range, while 
most are located in the green valley or blue sequence, which is consistent 
with the suggestion that most low-mass AGNs in GH07 reside in gas-rich, disk 
galaxies \citep{jiangyf11}.

We note that there still exist about 11\% of our sample objects in the red 
sequence. An interesting question arises: How were these LMBH AGNs triggered 
and fueled in such red and presumably gas-deficit galaxies? We know that AGN 
activity requires both available feeding fuel (gas or stars), and a process to get 
out of their angular momentum. As discussed in \citet{kormendy13}, the feeding 
rates of low-mass AGNs are very modest. Assuming these LMBHs accrete at 
their Eddington limits, the mass accretion rate $\dot{M} = 2.2 \times 10^{-8} 
(\eta/0.1) (M_{\rm BH}/M_{\odot}) M_{\odot}$, is only $\sim$0.02 \msun/year 
for BHs with $M_{\rm BH} \sim 10^6 M_{\odot}$ \citep {kormendy13}. These values 
are tiny even for red galaxies and hence there may always be enough gas to feed 
the nuclear low-mass AGNs. On the other hand, various physical process may 
break the ``angular momentum barrier'' and push the gas into the nuclear regions 
of galaxies such as wet major and minor mergers of the galaxies \citep[e.g.,][]
{silk98, hopkins08}, inflow along spiral arms or bar \citep[for instance, NGC 1097; 
e.g.,][]{davies09, fathi06} and disk instabilities \citep[e.g., cold flows;][]{dekel09, 
bournaud11}. Current observations show evidence that the triggering of 
low-mass AGNs are indeed dominated by secular processes. For instance, 
about 90\% of the LMBH AGNs in the low-redshift universe reside in galaxies with 
the so-called pseudobulges which are formed mainly by slow process without 
major mergers involved \citep{jiangyf11}. Our result on the 
$D_{4000}$--H$\delta_{\rm A}$ distribution also support this idea. The other 
processes, however, may not be completely ruled out. For example, SDSS 
J083803.68$+$540642.0, a red and gas-poor LINER in the sample of Dong+12, 
shows a peculiar circumgalactic ring, which may be caused by collision with a 
gas-rich galaxy, suggesting history of violent galaxy interaction \citep{liuwj17}. 
Such a process may also play a role in triggering low-mass AGN. As discussed 
in \citet{liuwj17}, major mergers of low-mass galaxies in the local universe may 
be more common than massive galaxies. Further studies are needed to fully 
understand this question. 

\subsection{Co-evolution of LMBHs and Their Host Galaxies?}
As discussed above, LMBHs seem to evolve with secular processes and have not 
experienced major mergers. Thus they can be used to trace the initial stage of 
the evolution of BHs and their host galaxies. We briefly investigate the co-evolution 
between BHs and host galaxies in the low-mass regime by using the relation 
between BH mass accretion rates and host star formation rates (SFRs). The 
mass accretion rates are derived from the bolometric luminosities assuming the 
efficiency factor $\eta=0.1$, while the SFRs are given by the MPA--JHU catalog 
with no aperture corrections. Figure~\ref{fig-sfr} shows the distribution of 282 
low-mass AGNs with both quantities available on the diagram of mass accretion 
rates versus SFRs, revealing a strong correlation between them. 

This suggests that there may exist a connection between the AGN activity and the 
star formation. However, more compelling and direct evidence is still needed to 
support the co-evolution of LMBHs and their hosts. In fact, there is no strong 
evidence that LMBHs have direct feedback on their host galaxies in the local 
universe. Furthermore, the link between BH accretion and star formation may 
also be explained by that they both depend on the gas from the same reservoir 
\citep{kormendy04}. Nevertheless, our result hints at a possible co-evolution 
scenario between LMBHs and their host galaxies.

\section{SUMMARY}
Using the optical spectrometric data from the SDSS DR7, we obtain a sample of 
204 new AGNs with low-mass black holes, and expand the SDSS LMBH sample 
from the DR4 (Dong+12) to a total of 513 objects. This is the largest optically 
selected, broad line low-mass AGN sample so far.  The BH masses, estimated 
using the virial method, are in the range of $1.1\times 10^5$ to $2.0 \times 
10^6$~\msun, with a median of $1.3 \times 10^6$~\msun. The Eddington ratios 
range from 0.01 to 2, with a median of 0.26. The properties and distributions 
of the new sample are statistically consistent with those of Dong+12.

We present some statistical properties of the combined LMBH AGN sample from 
this work and Dong+12, focusing on the emission lines and multi-wavelength 
properties including X-ray and radio, as well as their host galaxies. Most of the
LMBHs are located in the Seyfert galaxy and composite regions on the narrow-line 
diagnostic diagrams, confirming their AGN nature. 102 sources were detected by 
\rosat. LMBHs with X-ray detections tend to follow the correlation between the X-ray 
luminosities (2--10\,keV) and \oiii\,$\lambda$5007 luminosities derived from more 
massive AGNs. The optical/X-ray effective indices \aox\ of X-ray detected AGNs 
show a large scatter ranging from $-1.58$ to $-0.70$, and are systematically 
flatter than more massive AGNs. In general, they are broadly consistent with the 
extrapolation of the \aox--$L_{2500\angstrom}$ relation to the low-luminosity end.  
No dependence of \aox\ is found on \lratio, whereas a weak correlation between 
\aox\ and $M_{\rm BH}$ is suggested, which is consistent with \citet{dongrb12}. 
Only 5\% of the sources are detected in the FIRST survey, which are mostly radio 
loud. Thus we suggest that LMBHs are predominately radio-quiet, though further 
deep radio observations are needed to confirm this result. 

The host galaxies of LMBHs have $g$-band magnitudes ($M_g$) ranging from 
$-$22.2 to $-$15.9 mag, with a median comparable to the characteristic luminosity 
of $M^{*}_{g} = -$20.1~mag at $z = 0.1$. The colors of the galaxies suggest mostly 
a type of typical Sbc in general. The galaxies have stellar masses ($M_*$) 
ranging from $10^{8.8}$ to $10^{12.4}$~\msun, with a median of $10^{10.6}$~\msun, 
which is slightly lower than that of the SDSS DR7 sample from the MPA--JHU catalog.
Only a few tens have $M_* < 10^{10}$ \msun\ or $M_g > -18$~mag. Thus low-mass 
BHs may live in lower-mass stellar systems, but do not necessarily reside in the 
dwarf galaxies. Most of the galaxies have mean stellar ages younger than 1~Gyr from 
their $D_{4000}$ values. The locus on the $D_{4000}$--H$\delta_{\rm A}$ diagram 
indicates that they tend to have experienced continuous star formation over the past 
few Gyrs, which is consistent with the suggestion that their host galaxies of LMBHs mainly 
evolve via secular processes \citep[e.g.,][]{jiangyf11, kormendy13}. Their distribution 
on the color versus $M_*$ diagram shows that most of these are blue, late-type galaxies.

With homogeneous selection and accurately measurements of the spectral 
parameters, our SDSS LMBH sample provides a useful database to further 
explore the properties of low-mass BHs and their host galaxies, as well as 
to study the BH mass function in the low-mass regime.

\acknowledgments
This work is supported by the National Natural Science Foundation of China 
(grant No.\,11473035 and No.\,11473062) and the National Program on Key 
Research and Development Project (Grant No.\,2016YFA0400804). 
W.L. acknowledges supports from the Natural Science Foundation of China 
grant (NSFC 11703079) and the ``Light of West China'' Program of Chinese 
Academy of Sciences (CAS). H.L. thanks NAOC for providing the computing 
resources on Zen cluster. We are grateful to the anonymous referee for his/her 
constructive comments that improved the paper. This work is mainly based on 
the observations obtained by the SDSS, we acknowledge the entire SDSS 
team for providing the data that made this work possible. We have made use 
of the \rosat\  Data Archive of the Max-Planck-Institut f{\"u}r extraterrestrische 
Physik (MPE) at Garching, Germany. This publication also makes use of data 
products from the Wide-field Infrared Survey Explorer (WISE) and the Two 
Micron All Sky Survey (2MASS). WISE is a joint project of the University of 
California, Los Angeles, and the Jet Propulsion Laboratory/California Institute 
of Technology, which are funded by the National Aeronautics and Space 
Administration (NASA). 2MASS is a joint project of the University of Massachusetts 
and the Infrared Processing and Analysis Center/California Institute of Technology, 
funded by the NASA and the National Science Foundation (NSF). We also use the 
data from the UKIDSS, which uses the UKIRT Wide Field Camera. 

\bibliography{liuheyang}
\clearpage


\setcounter{figure}{0}
\setcounter{table}{0}
\begin{figure}[bp]
\epsscale{1} \plotone{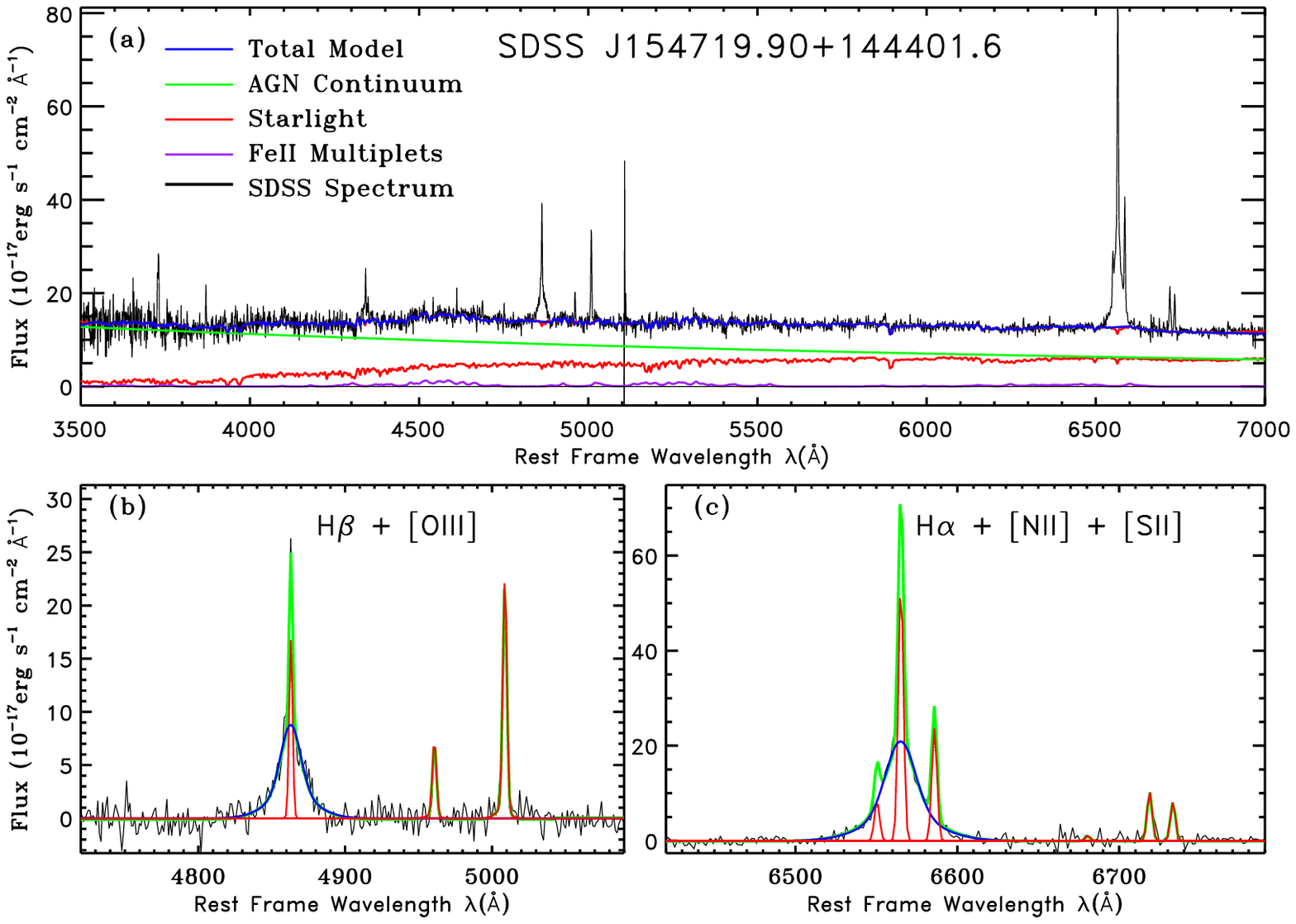}
\caption{\label{fig-illustration}%
Illustration of the continuum and emission-line fitting for one of our LMBH 
AGNs as an example. %
Panel ($a$): The observed SDSS spectrum (black), the total model
(blue), the decomposed components of the host galaxy (red), the AGN
continuum (green), and the \feii\ multiplets (purple).
Panel ($b$): Emission-line profile fitting in the \hb~$+$~\oiii\ region.
Panel ($c$): Emission-line profile fitting in the \ha~$+$~\nii~$+$~\sii\ region.
({\it The illustration of the continuum and emission-line fitting for all the 
LMBH AGNs are available in the online journal.})
}
\end{figure}

\begin{figure}[tbp]
\epsscale{1} \plotone{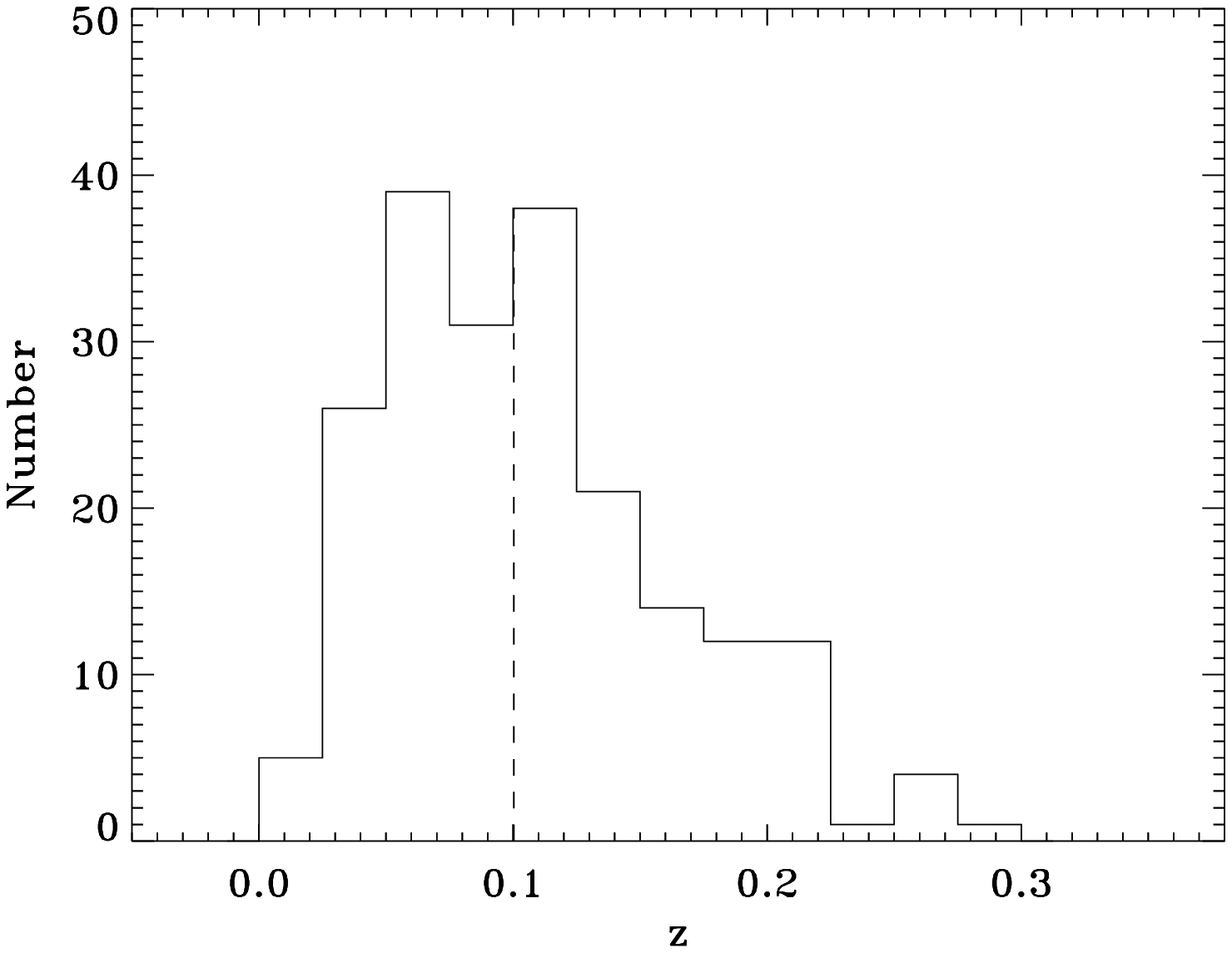} %
\caption{\label{fig-redshift}%
Redshift distribution for our 204 low-mass AGNs. The
dashed line denotes the median.}
\end{figure}

\begin{figure}[tbp]
\centering
\begin{minipage}[]{1.0\hsize}
\includegraphics[width=\hsize,angle=0]{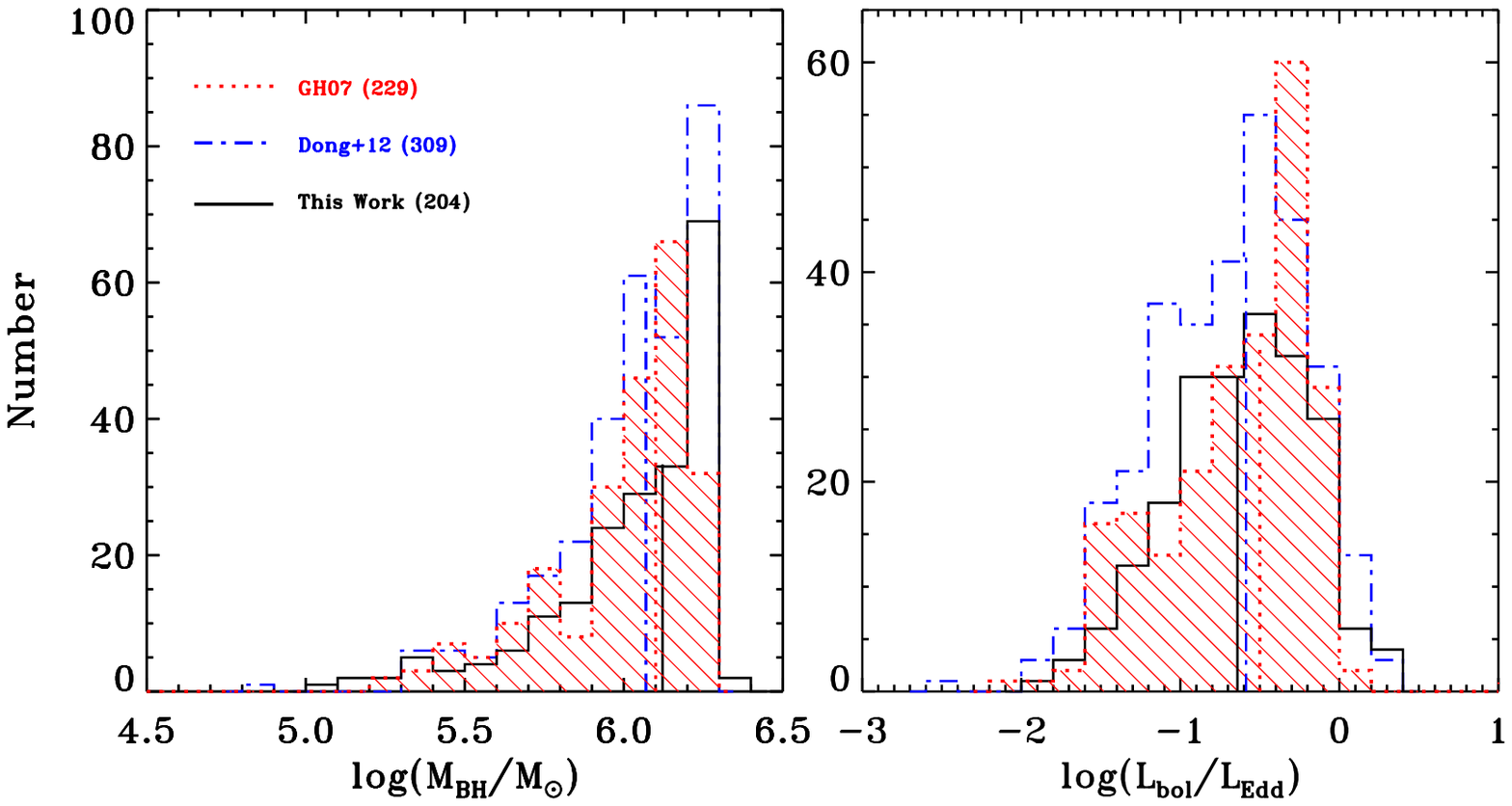}
\vspace{-4cm}
\end{minipage}
\begin{minipage}[]{1.0\hsize}
\includegraphics[width=\hsize,angle=0]{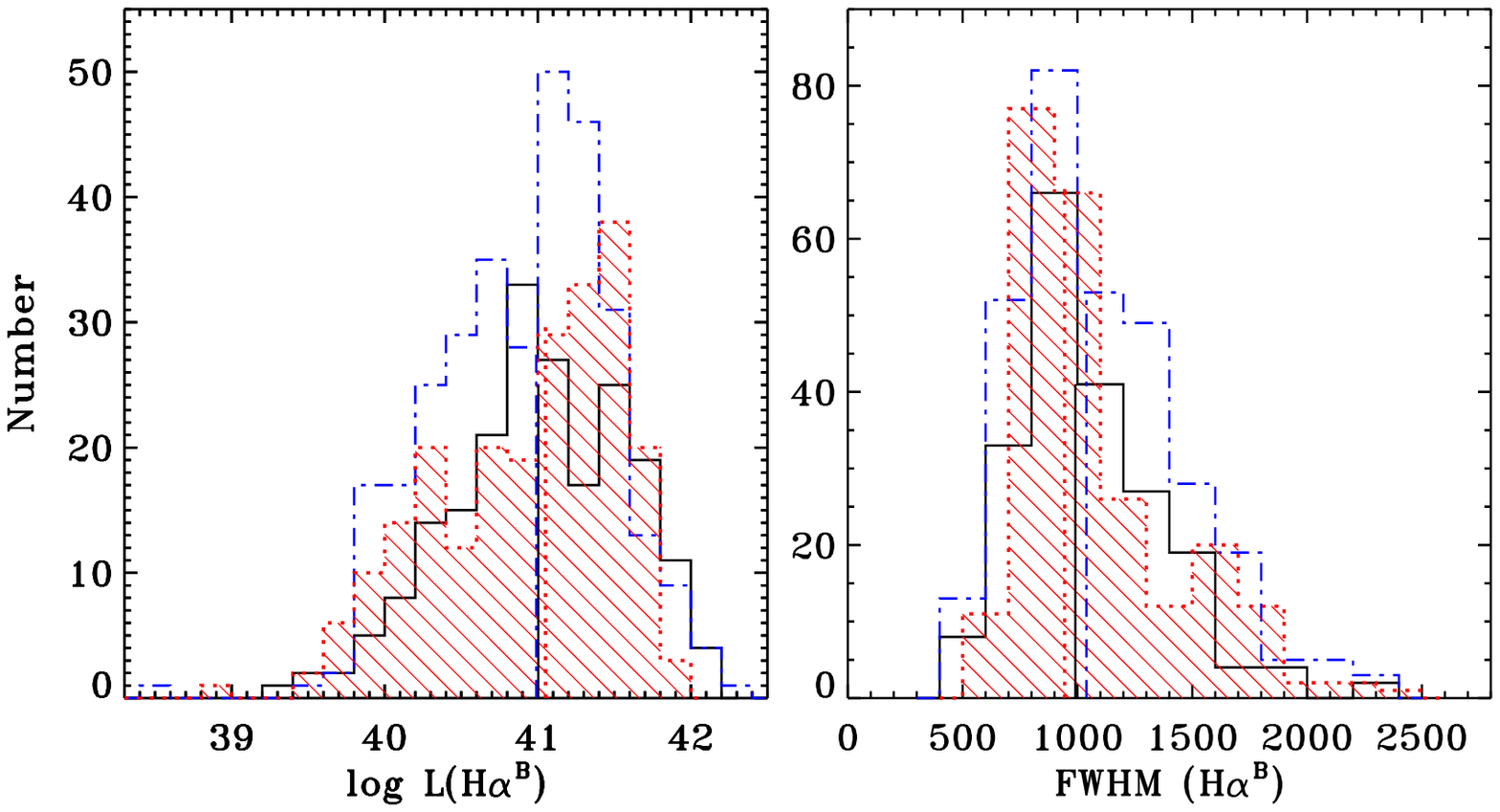}
\end{minipage}
\caption{\label{fig-ourdist}
Distributions of the BH mass, Eddington ratio, luminosity and the FWHM of
broad \ha\,for the LMBH sample in this work (black solid histograms), GH07 
(red dotted histograms) and Dong+12 (blue dotted dashed histograms).
The vertical lines denote the corresponding medians. Numbers in 
brackets of the upper left panel indicate the sample sizes. 
}
\end{figure}

\begin{figure}[tbp]
\epsscale{1} \plotone{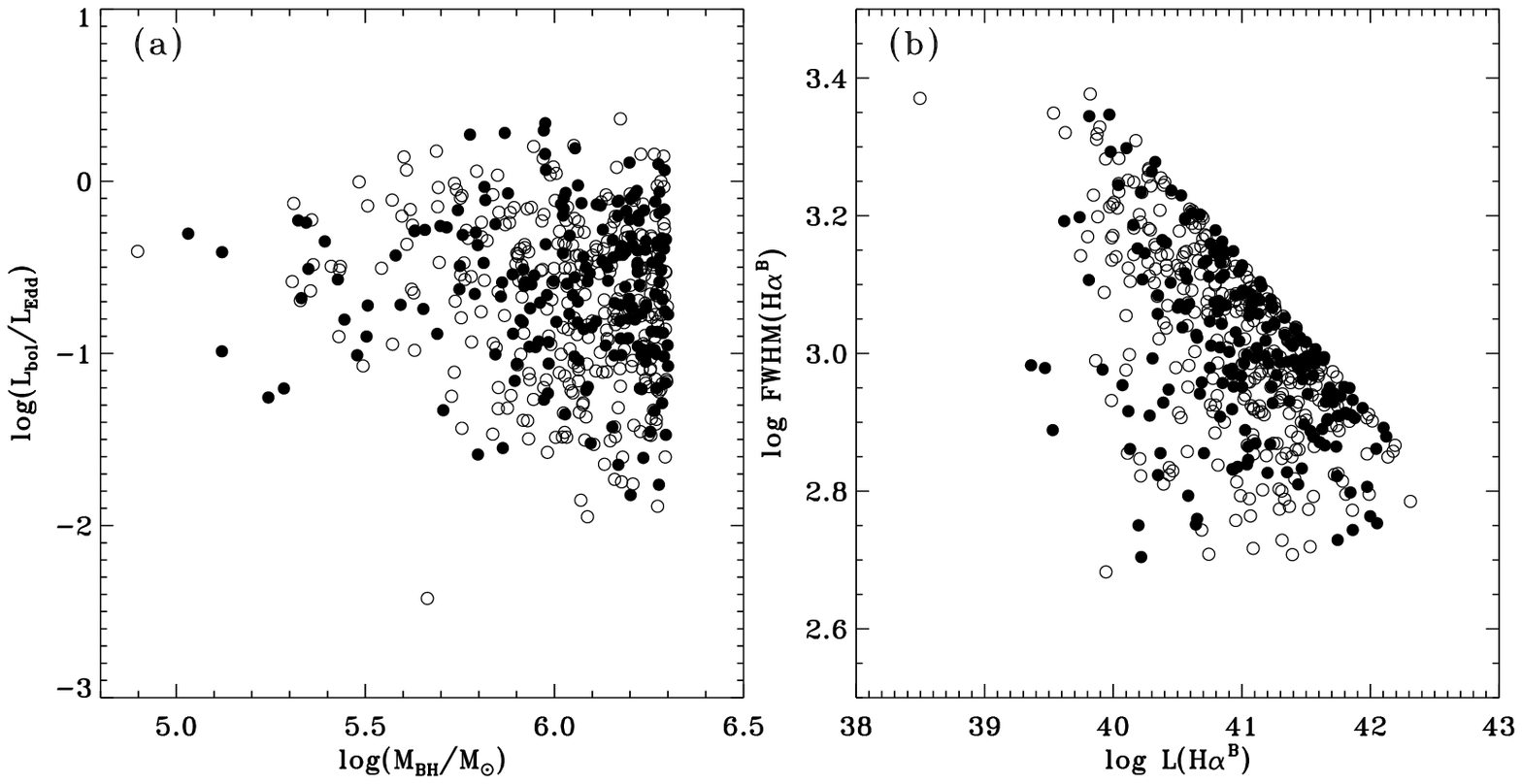} %
\caption{\label{fig-optpara}%
Distributions of the LMBH sample in this work (filled circles) and Dong+12 
(open circles) on the \lratio\ versus $M_{\rm BH}$ plane ({\it a}) and on the 
\fwbha\ versus \lbha\ plane ({\it b}).}
\end{figure}

\begin{figure}[tbp]
\epsscale{1} \plotone{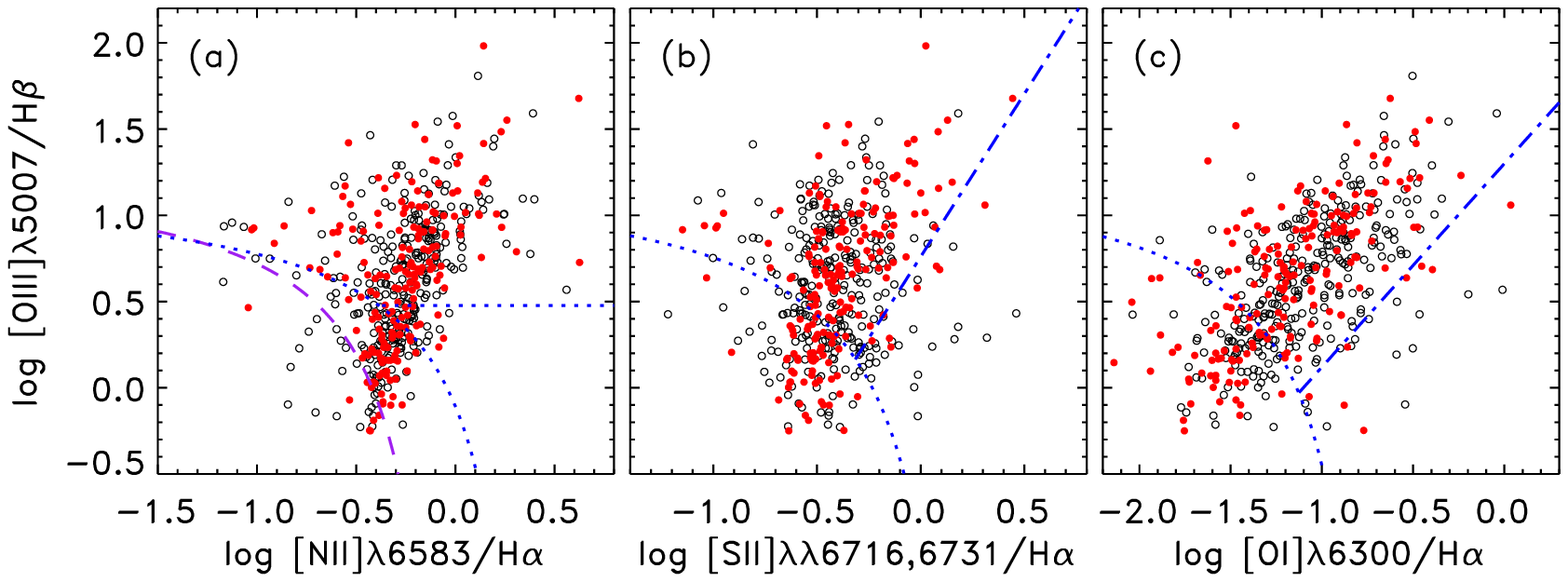} %
\caption{\label{fig-bpt}%
Narrow-line diagnostic diagrams of \oiii\,$\lambda 5007/$\hb\ versus
\nii\,$\lambda 6583/$\ha\ ({\it a}), versus \sii\,$\lambda\lambda 
6716,6731/$\ha\ ({\it b}), and versus \oi\,$\lambda 6300/$\ha\ ({\it c}) 
for the LMBH sample in this study (black open circles) and Dong+12 
sample (red filled circles). The extreme starburst classification line 
(blue dotted curve) from \citet{kewley01} and the Seyfert--LINER line
(blue dotted dashed line) obtained by \citet{kewley06} are adopted
to separate \hii\ regions, AGNs and LINERs. In panel ($a$), the purple
dashed line corresponds to the pure star formation line given by
\citet{kauffmann03b}, and the blue dotted horizontal line represents
\oiii\,$\lambda 5007/$\hb\ $=3$, which is conventionally used to 
separate Seyfert galaxies and LINERs.}
\end{figure}


\begin{figure}[tbp]
\epsscale{1} \plotone{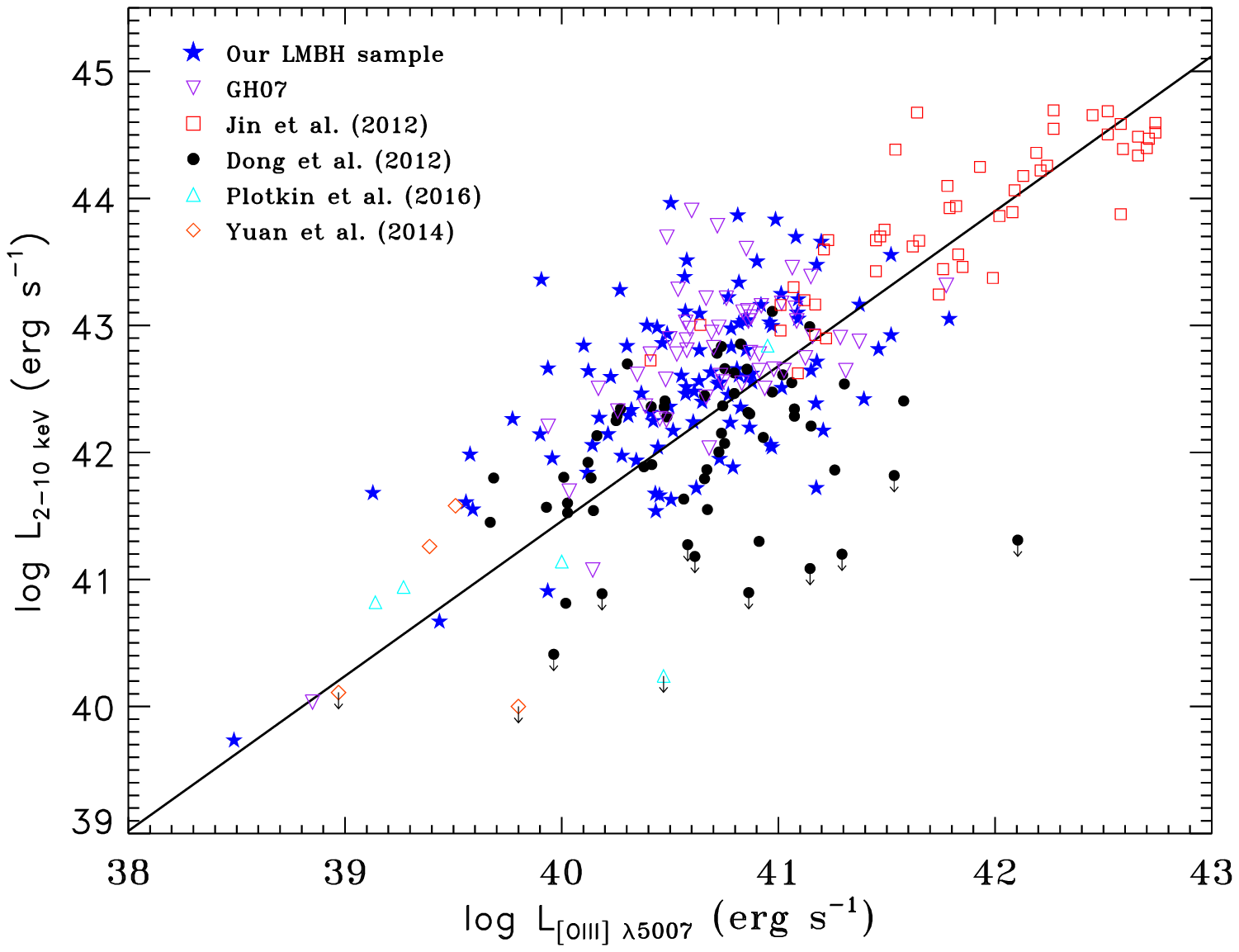} %
\caption{\label{fig-optLx}%
X-ray luminosity in 2--10 keV versus \oiii\,$\lambda5007$ luminosity 
for our total LMBH sample detected with \rosat\ (blue filled star symbols), 
GH07 sample with \rosat\ detections (purple open inverted triangles),
LMBHs with \chandra\ detections in \citet[black filled circles]{dongrb12}, 
low-mass active galaxies with low Eddington ratios from \citet[orange red 
open diamonds]{yuanwm14} and \citet[cyan open triangles]{plotkin16}, 
and more massive AGNs from \citet[red open squares]{jincc12}. The 
black solid line represents the relation for Seyfert galaxies and QSOs 
given by \citet{panessa06}. Arrows denote upper limits.
}
\end{figure}

\begin{figure}[tbp]
\epsscale{1} \plotone{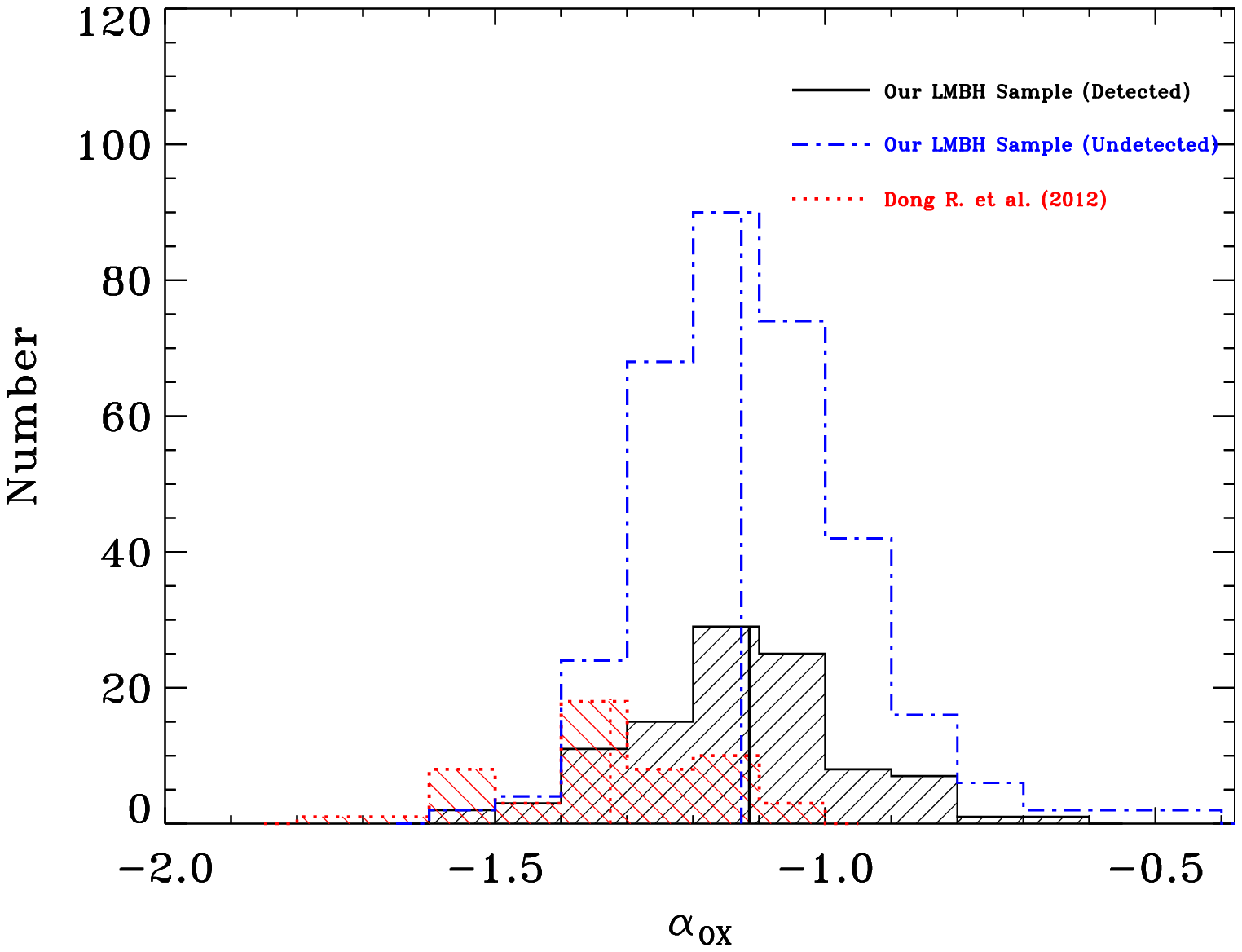} %
\caption{\label{fig-aoxdist}%
Distributions of \aox\ for objects detected with \rosat\ in the total LMBH 
sample (black shaded histogram). \chandra-detected sources in 
\citet[red dotted shaded histogram]{dongrb12} and upper limits for 
those undetected in \rosat\ (blue dotted dashed histogram) in our sample 
are also plotted for comparison. The vertical lines represent the 
corresponding medians.
}
\end{figure}

\begin{figure}[tbp]
\epsscale{1} \plotone{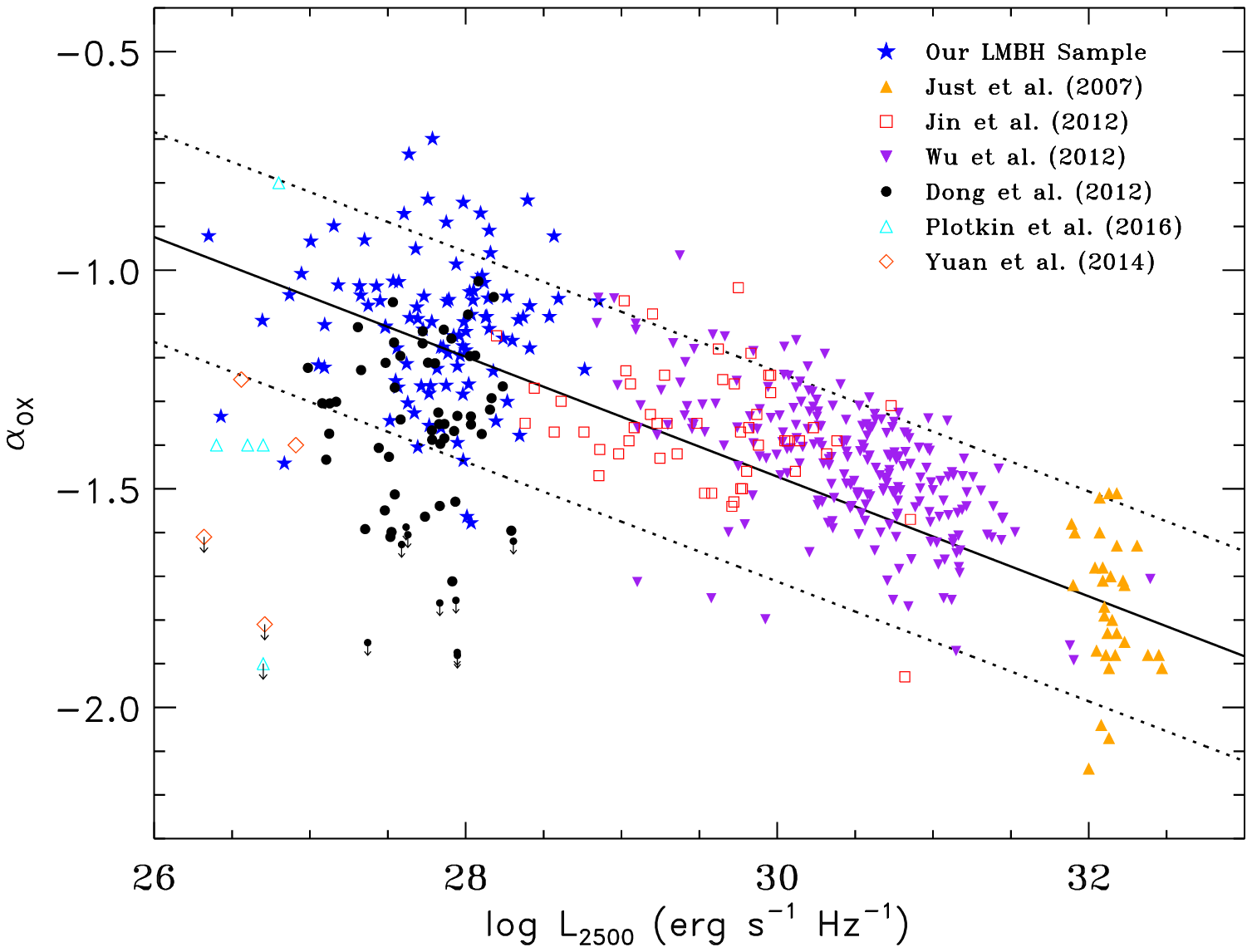} %
\caption{\label{fig-aoxl2500}%
Optical-to-X-ray spectral index \aox\ versus the monochromatic luminosity 
at 2500~\AA\ for the total LMBH sample detected with \rosat\ (blue filled star 
symbols). The 2500~\AA\ monochromatic luminosities are derived from 
\lbha\ using the scaling relation between \lbha\ and \lfive\ assuming a 
spectral shape of 1.56 ($f_{\lambda} \propto \lambda^{-1.56}$). LMBHs in 
\citet[black filled circles]{dongrb12}, \citet[orange red open diamonds]
{yuanwm14} and \citet[cyan open triangles]{plotkin16}, as well as more 
luminous Seyfert galaxies and QSOs in \citet[red open squares]{jincc12}, 
\citet[purple filled inverted triangles]{wujian12} and \citet[orange filled 
triangles]{just07} are also plotted for comparison. The solid and dashed 
lines represent the relation and 1~$\sigma$ scatter given by \citet{steffen06},
respectively. Arrows represent upper limits.
}
\end{figure}

\begin{figure}[tbp]
\epsscale{1} \plotone{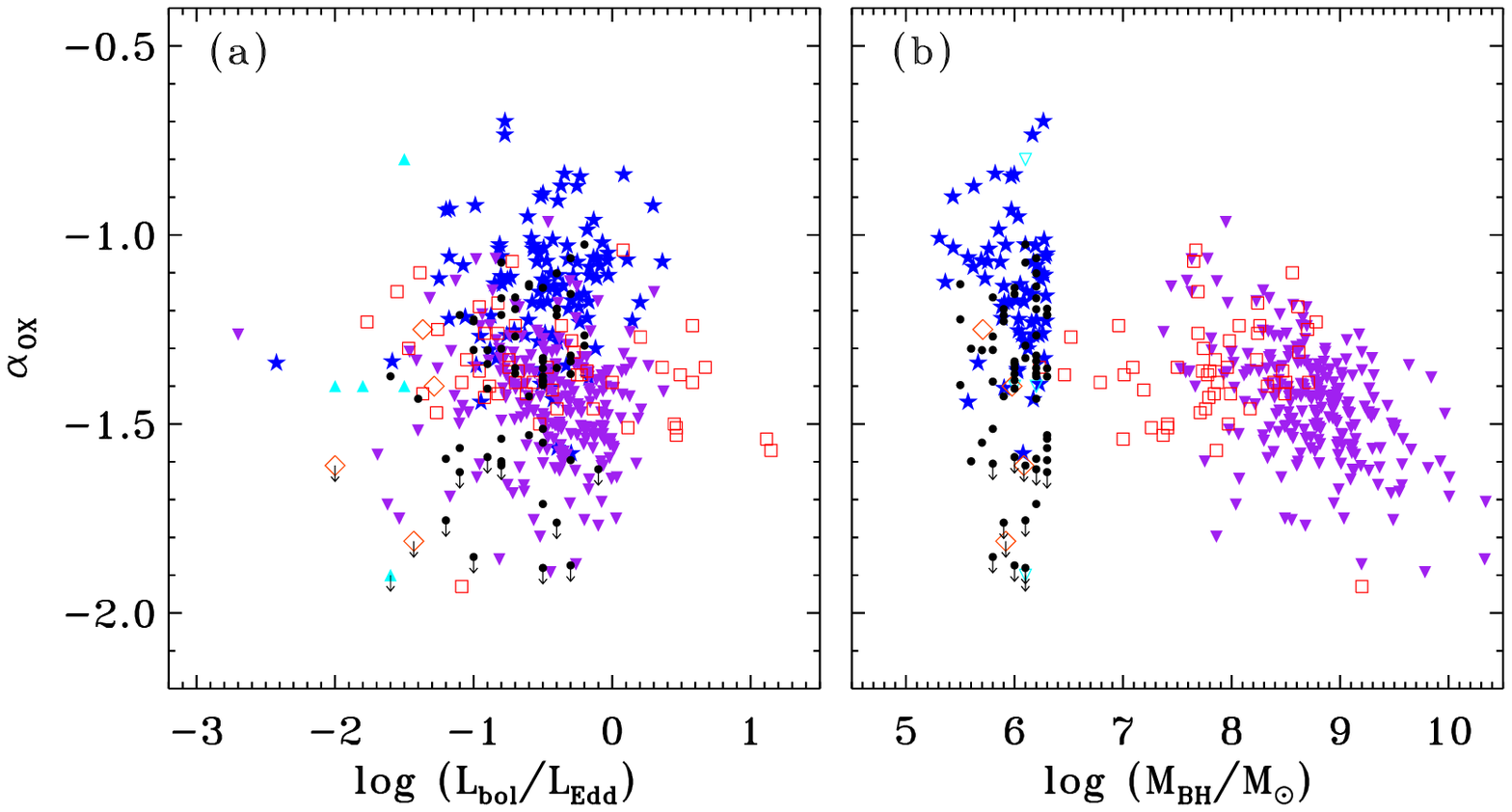} %
\caption{\label{fig-aoxmbh}%
Dependence of the optical-to-X-ray spectral index \aox\ on Eddington ratio 
({\it a}) and BH mass ({\it b}). The blue filled stars represent the total LMBH 
sample. Overplotted symbols are the low-mass AGNs with the Eddington 
ratios log $(L_{\rm bol}/L_{\rm Edd}) < -1$ from \citet[orange red open 
diamonds]{yuanwm14} and \citet[cyan open triangles]{plotkin16}, LMBHs 
observed by \chandra\ from \citet[black filled circles]{dongrb12}, and more
massive AGNs from \citet[purple filled inverted triangles]{wujian12} and 
\citet[red open squares]{jincc12}, respectively. Arrows denote upper limits 
on \aox.
}
\end{figure}


\begin{figure}[tbp]
\epsscale{1} \plotone{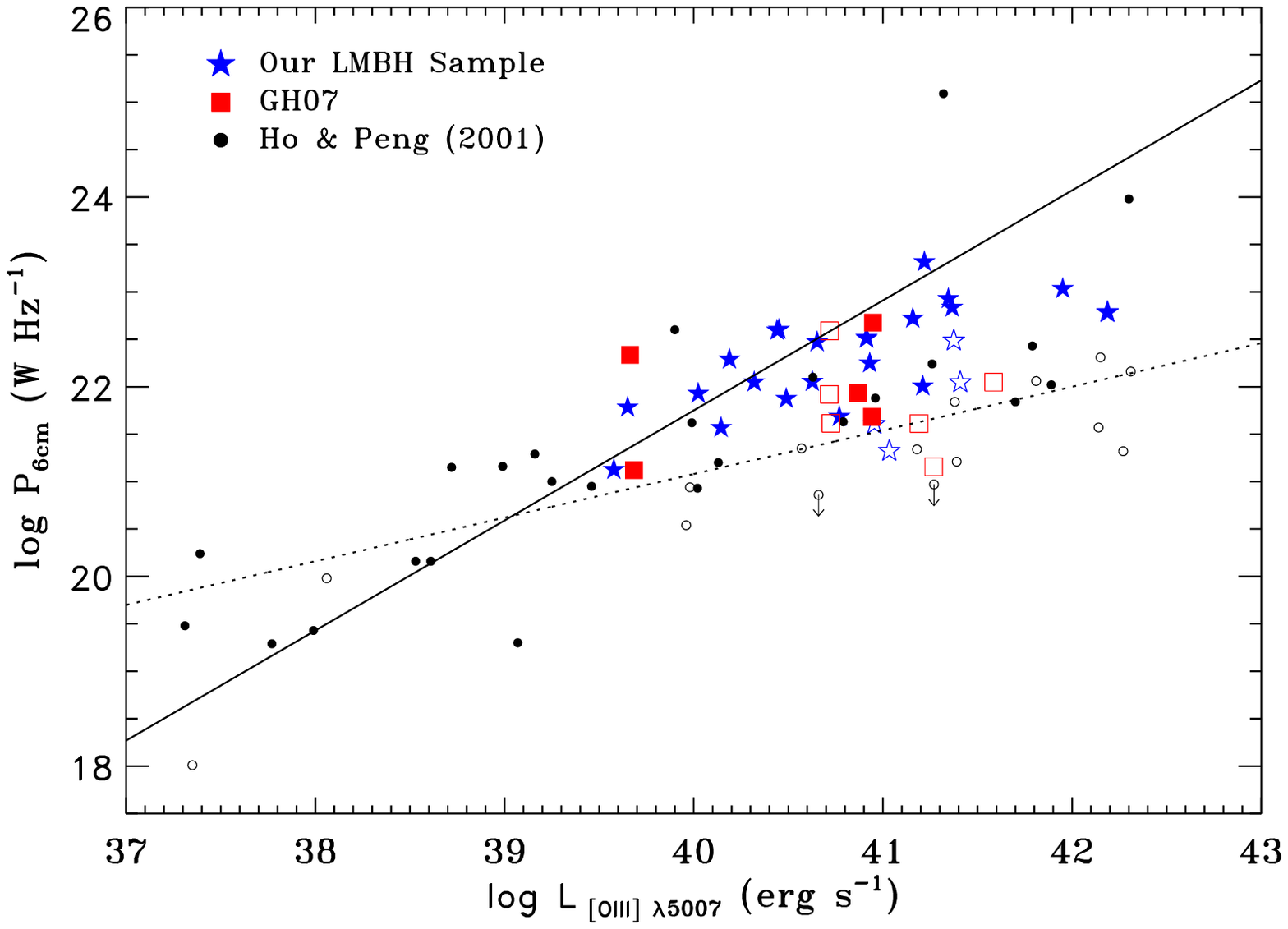} %
\caption{\label{fig-radiop6cm}%
Relation of the radio power at 6~cm and the \oiii\,$\lambda5007$ luminosity 
for our total LMBH sample (blue stars), low-mass black holes from GH07 
(red squares) and more massive Seyfert galaxies from \citet[black circles]
{ho01}. The filled and open symbols represent radio-loud and radio-quiet 
sources, respectively, and arrows denote upper limits. The solid and dashed
lines represent the relation derived from more massive radio-loud and 
radio-quiet AGNs, respectively \citep{ho01}. 
}
\end{figure}

\begin{figure}[tbp]
\epsscale{1} \plotone{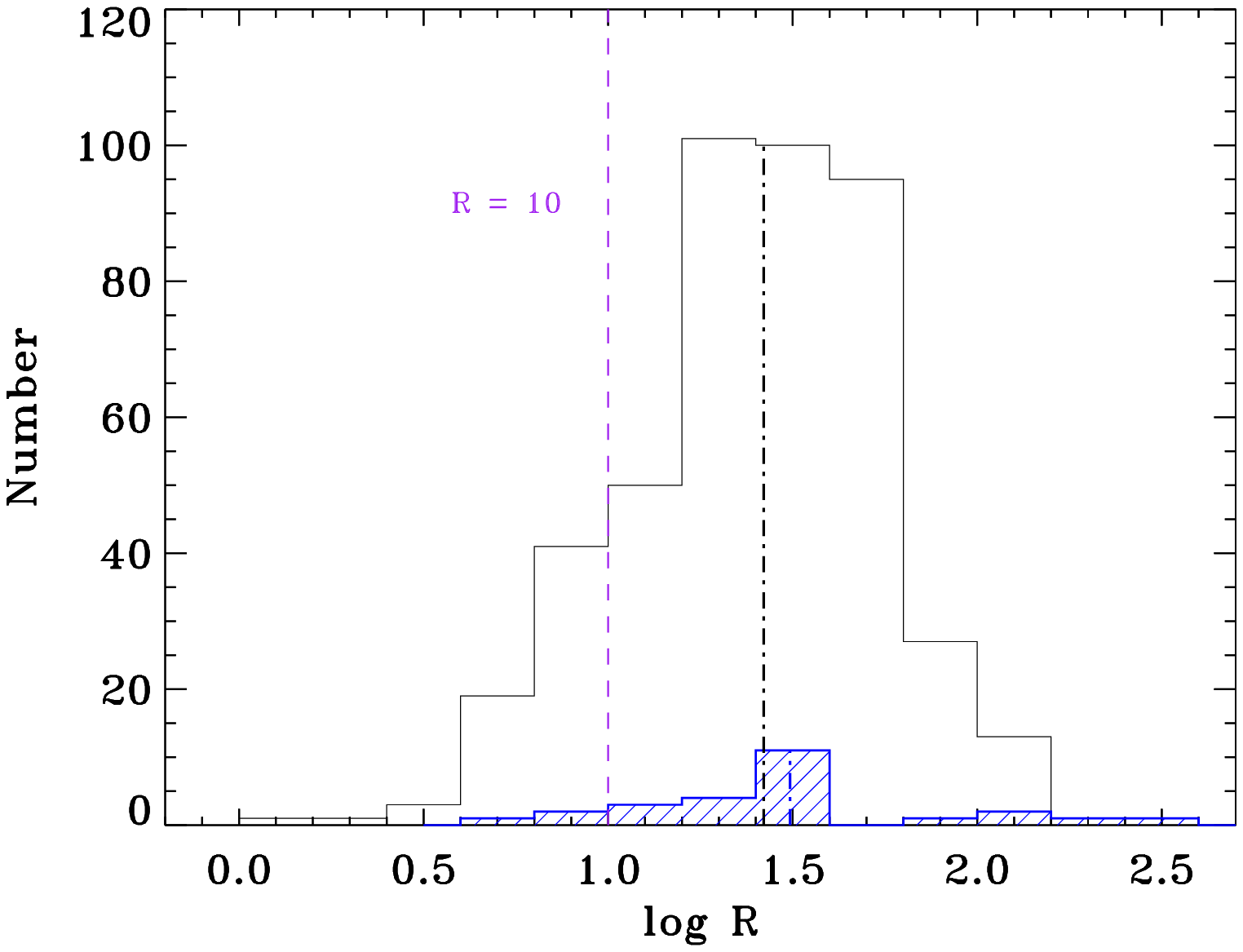} %
\caption{\label{fig-radioloudness}%
Distribution of the radio loudness of the total LMBH sample objects. Blue 
shaded histogram represents sources detected in the FIRST, while 
the black histogram represents upper limits derived assuming a flux density 
limit of 1~mJy at 20~cm for the FIRST survey. The conventional 
demarcation line between radio-loud and radio-quiet AGNs is
marked by the dashed line. The dotted dashed lines denote 
corresponding medians.}
\end{figure}

\begin{figure}[tbp]
\epsscale{1} \plotone{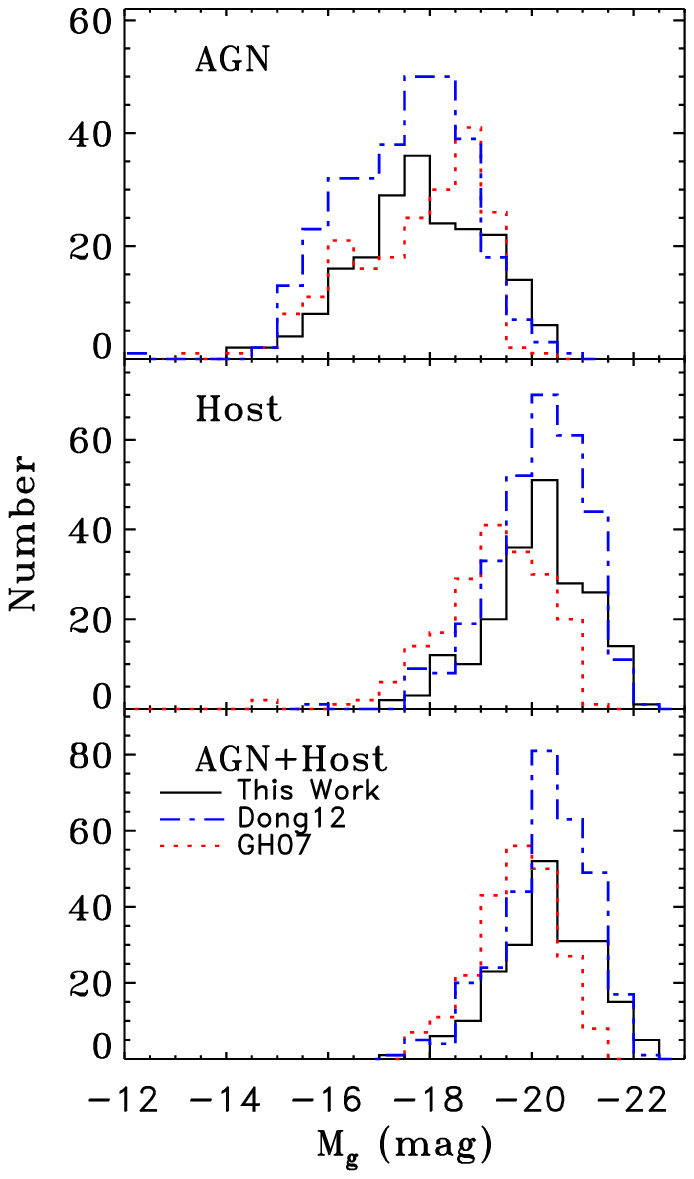} %
\caption{\label{fig-mag}%
Distributions of the absolute $g$-band magnitudes of AGN ({\it top}), 
host galaxy ({\it middle}), and the total (AGN plus host galaxy) ({\it 
bottom}), for the sample in this work (black solid line), Dong+12 
(blue dotted dashed line) and GH07 (red dotted line), respectively
(see text for details of the estimation of the AGN and host galaxy 
luminosities.) 
}
\end{figure}

\begin{figure}[tbp]
\epsscale{1} \plotone{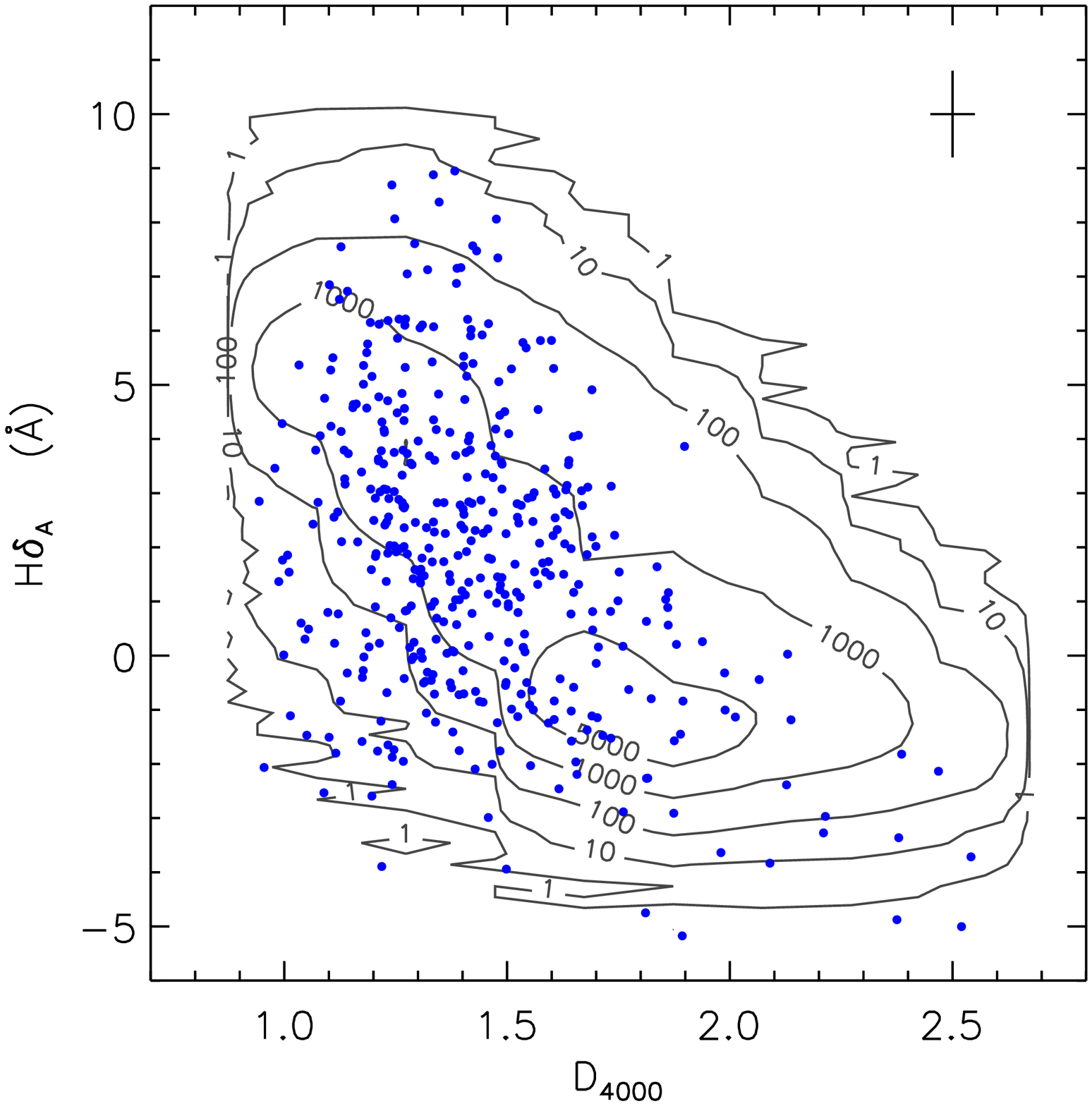} %
\caption{\label{fig-d4k}%
Distribution of 423 low-mass BH host galaxies (blue filled circles) from
the total LMBH sample on the plane of 4000 \AA\,break strength 
($D_{4000}$) versus equivalent width of the H$\delta$ absorption 
(H$\delta_{\rm A}$). As a comparison, the distribution of $\sim$492,000 
non-AGN galaxies in the SDSS DR7 is also plotted as contours.
The cross at the upper-right corner represents the
typical size of 1~$\sigma$ errors.}
\end{figure}

\begin{figure}[tbp]
\epsscale{1} \plotone{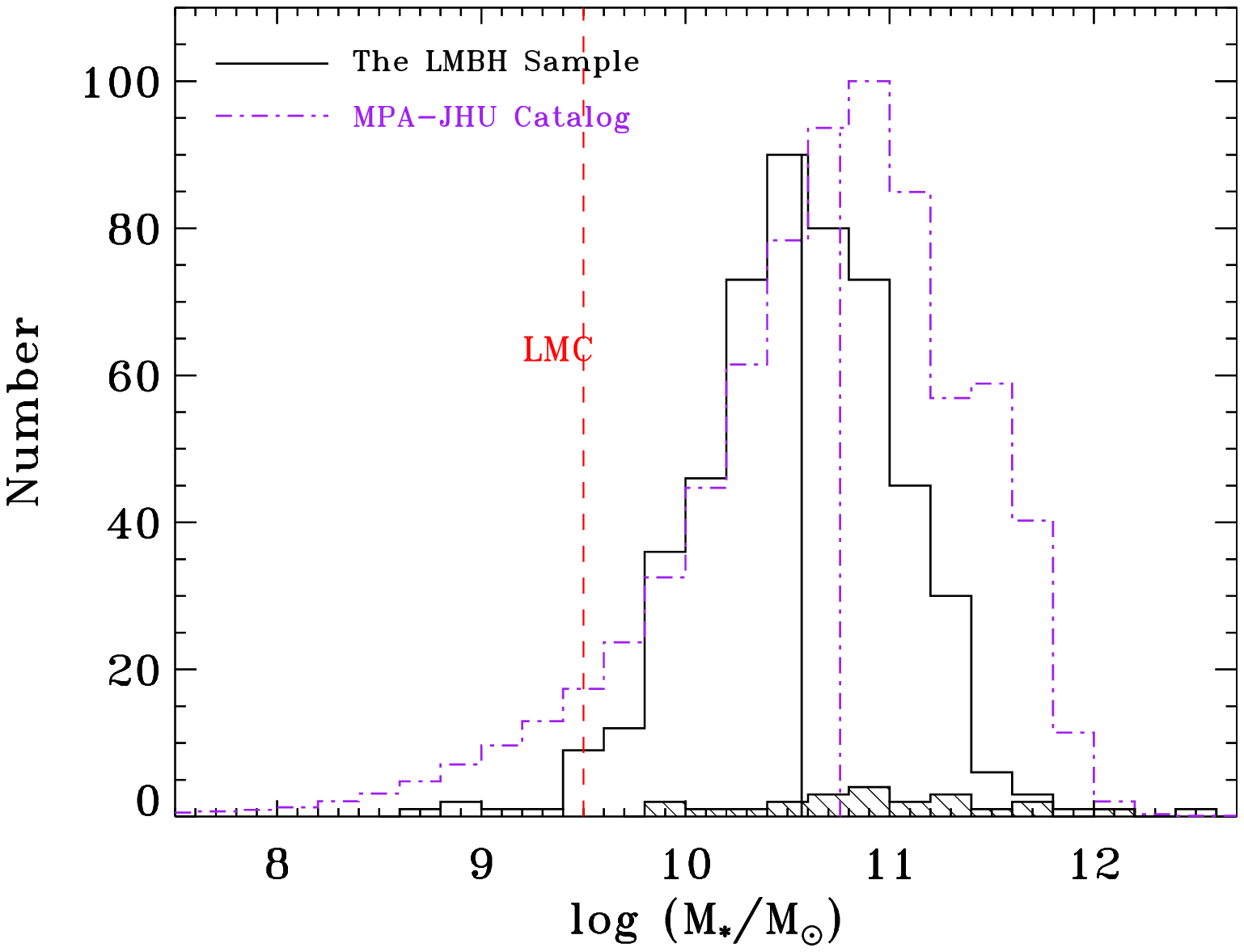} %
\caption{\label{fig-stellarmassdist}%
Distributions of the stellar masses ($M_*$) of the host galaxies of our 
LMBH sample (black solid histogram) and $\sim$925000 sources in 
the SDSS DR7 of which  the stellar masses are taken from the 
MPA--JHU catalog (purple dotted dashed histograms). The vertical lines 
represent the corresponding medians. The black shaded histogram 
represents radio-loud AGNs in our LMBH sample. (The MPA--JHU 
histogram is normalized to have a peak value of 100 for ease of 
comparison.)
}
\end{figure}

\begin{figure}[tbp]
\epsscale{1} \plotone{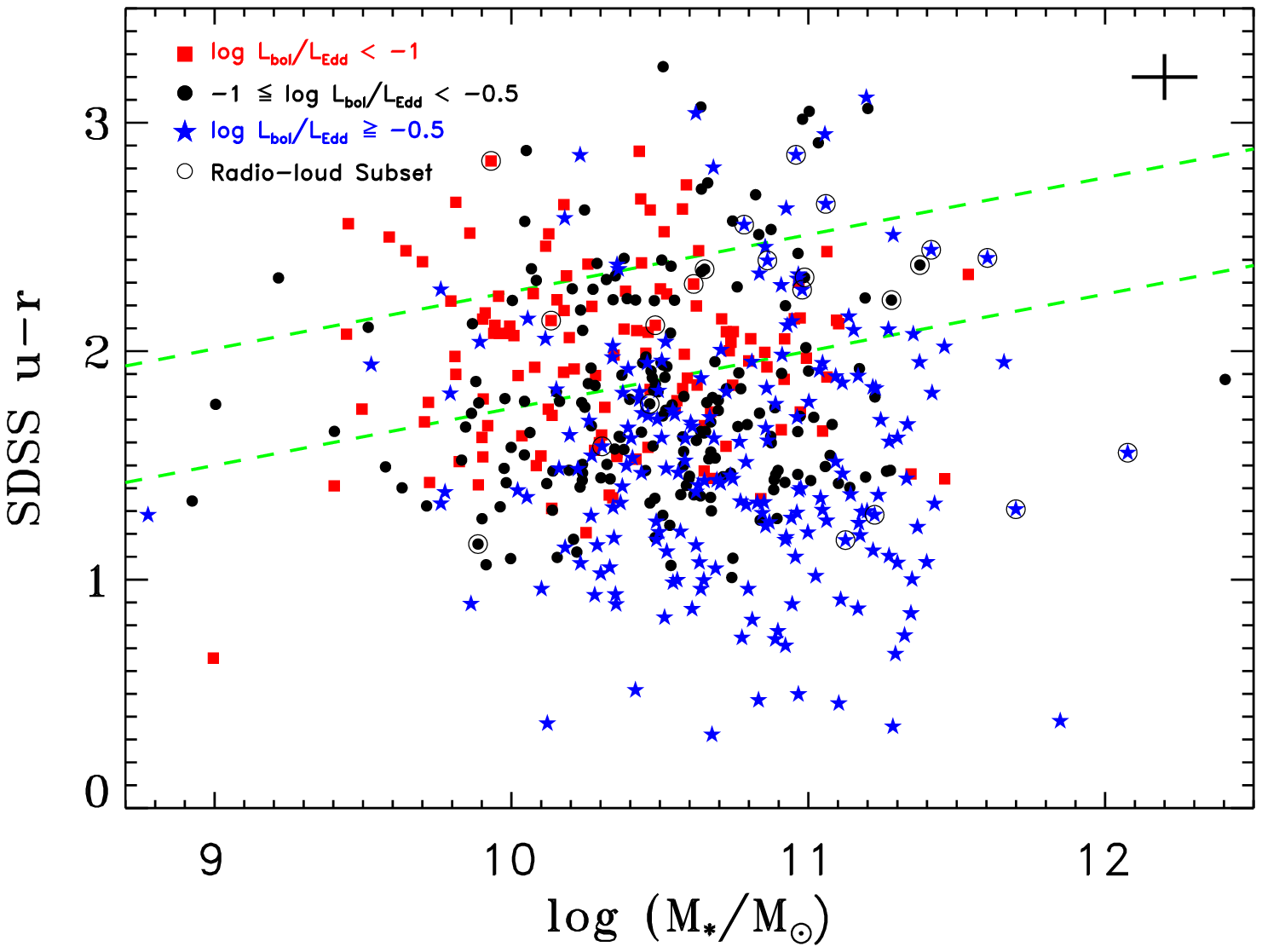}
\caption{\label{fig-stellarmasscolor}%
Distribution of the total LMBH sample on the the $u - r$ color 
versus stellar mass plane. The green dashed lines denote the green 
valley defined by \citet{schawinski14}. Red square, black filled circle 
and blue star represent sources with their Eddington ratios in the 
ranges of log \lratio\,$< -1$, $-1\leq$~log \lratio~$<-$0.5 and log 
\lratio~$\geq$~$-$0.5, respectively. Open circles denote radio-loud 
sources. Our LMBH sample objects span almost the entire $u - r$ 
color range, and most of the objects are located in the green valley 
or blue sequence, while about 11\% of the sample are in the red 
sequence. The cross at the upper right corner represents the typical 
size of 1~$\sigma$ errors.
}
\end{figure}

\begin{figure}[tbp]
\epsscale{1} \plotone{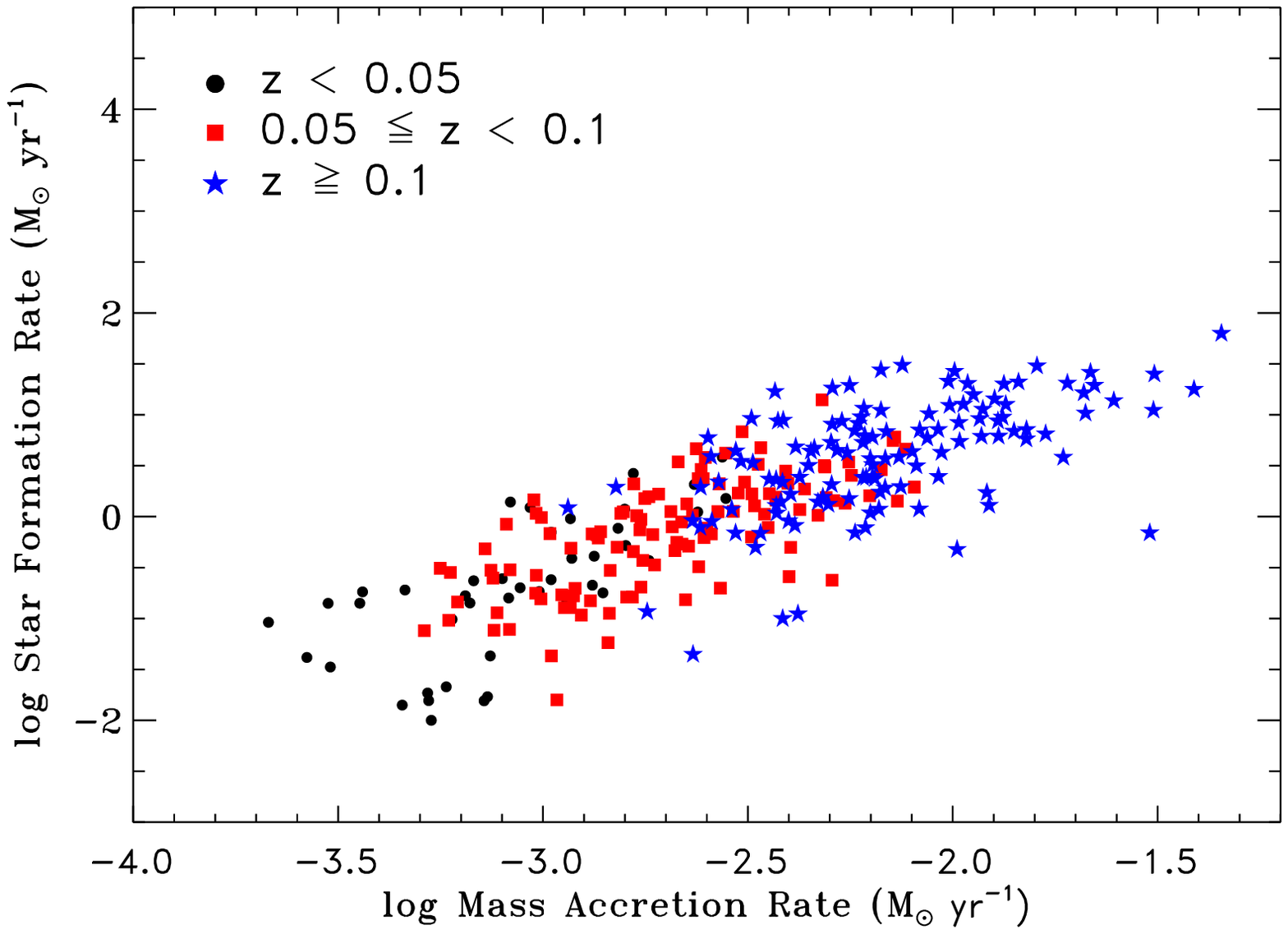}
\caption{\label{fig-sfr}%
Mass accretion rate versus star formation rate for 282 sources in the total 
LMBH sample with AGN contribution lower than 75\%. The mass accretion 
rates are derived assuming an efficiency of $\eta$\,=\,0.1. The star 
formation rates are obtained from the MPA--JHU catalog estimated using 
a technique described in \citet{brinchmann04}. Blue stars, red squares 
and black circles represent sources with redshift $z \geq 0.1$, $0.05 
\leq z < 0.1$ and $z < 0.05$ respectively.
}
\end{figure}

\clearpage


\begin{deluxetable}{rclccc}
\tablecolumns{6} \tabletypesize{\scriptsize} \tablewidth{0pc}
\tablecaption{The SDSS DR7 LMBH Sample} %
\tablehead{ \colhead{ID}  & \colhead{SDSS Name} & \colhead{$z$} &
\colhead{$g$} & \colhead{$g-r$} & \colhead{$A_{g}$}
\\
\colhead{(1)} & \colhead{(2)} & \colhead{(3)} &
\colhead{(4)} & \colhead{(5)} & \colhead{(6)}
}
\startdata
 1     &    J054248.74$+$004019.2  &   0.0518   &    17.49  &   0.98   &  1.14    \\  
 2     &    J074345.47$+$480813.5  &   0.0181   &    16.24  &   0.74   &  0.22    \\  
 3     &    J080807.13$+$563832.4  &   0.0990   &    18.93  &   0.55   &  0.18        
\enddata
\tablecomments{
Col. (1): Identification number assigned in this paper.
Col. (2): Official SDSS name in J2000.0.
Col. (3): Redshift measured by the SDSS pipeline.
Col. (4): Petrosian $g$ magnitude, uncorrected for Galactic extinction.
Col. (5): Petrosian $g-r$ color.
Col. (6): Galactic extinction in the $g$ band.
{\it (This table is available in its entirety in a machine-readable form in the online
journal. A portion is shown here for guidance regarding its form and content.)}
}
\label{tab-basic}
\end{deluxetable}

\begin{sidewaystable*}[htbp]
\topmargin 0cm \evensidemargin = 0mm \oddsidemargin = 0mm
\scriptsize

\caption{Emission-line Measurements}
\centering
\vspace{0.5cm}
\begin{tabular}{rcccccccccccccc}
\hline \hline
ID &
[O\,{\tiny II}] $\lambda3727$ &
Fe\,{\tiny II} $\lambda4570$  &
H$\beta^{\rm N}$ &
H$\beta^{\rm B}$ &
[O\,{\tiny III}] $\lambda5007$ &
[O\,{\tiny I}] $\lambda6300$   &
H$\alpha^{\rm N}$ &
H$\alpha^{\rm B}$ &
[N\,{\tiny II}] $\lambda6583$ &
[S\,{\tiny II}] $\lambda6716$ &
[S\,{\tiny II}] $\lambda6731$ &
FWHM$_{\rm H\alpha^B}$  &
FWHM$_{\rm [O\,III]}$   &
FWHM$_{\rm [S\,II]}$    \\
(1) & (2)  & (3)  & (4)  & (5)  & (6)  & (7) & (8) &
(9) & (10) & (11) & (12) & (13) & (14) & (15)  \\
\hline
 1   &  $<-$15.46  &  $<-$14.05  &    $-$15.57  &  $<-$15.54  &   $-$14.98  &    $-$15.73  &   $-$14.82  &   $-$14.70 &  $-$15.02  &   $-$15.59  &   $-$15.64  &   1986   &     227   &      133  \\   
 2   &    \nodata  &  $<-$14.54  &    $-$15.12  &   $-$14.92  &   $-$14.27  &    $-$15.16  &   $-$14.43  &   $-$13.95 &  $-$14.91  &   $-$15.00  &   $-$15.06  &    947   &     222   &      166  \\   
 3   &   $-$14.80  &   $-$15.08  &    $-$14.99  &   $-$14.80  &   $-$14.38  &    $-$15.59  &   $-$14.39  &   $-$14.21 &  $-$14.62  &   $-$15.08  &   $-$15.20  &   1043   &     204   &      136  \\   
\hline
\end{tabular}
\medskip
\vfill
{\normalsize Note. ---
Col. (1): Identification number assigned in this paper. %
Cols. (2)--(12): Emission-line fluxes (or 3 $\sigma$ upper limits) in
log-scale, in units of $\mathrm{erg~s^{-1}~cm^{-2}}$. The measured
emission-line fluxes are regarded to be reliable detections if they
have significance greater than 3~$\sigma$, or else the 3~$\sigma$ values
will be adopted as the upper limits. Note that
these are observed values only corrected Galactic extinction, and
no NLR or BLR extinction correction has been applied. The superscripts
``N'' and ``B'' in cols. 4, 5, 8, 9 and 13 refer to the narrow and broad
components of the line, respectively. %
Cols. (13)--(15): Line widths (FWHM) that are calculated from the
best-fit models and have been corrected for instrumental broadening
using the values measured from arc spectra and tabulated by the
SDSS; in units of \kms. %
{\it (This table is available in its entirety in a machine-readable
form in the online journal. A portion is shown here for guidance
regarding its form and content.)} }
\label{tab-emline}
\end{sidewaystable*}

\begin{deluxetable}{rccccccccc}
\tablecolumns{10} \tabletypesize{\scriptsize} \tablewidth{0pc}
\tablecaption{Luminosity, Mass and Eddington Ratio Measurements} %
\tablehead{ \colhead{ID}  & \colhead{$M_{g}$(total)} & \colhead{$M_{g}$(AGN)} &
\colhead{$M_{g}$(host)} & \colhead{log $L_{\rm H\alpha^B}$} & \colhead{log $L_{\rm [O\,{\tiny III}] \lambda5007}$} &
\colhead{log $M_{\rm BH}$} & \colhead{log $L_{\rm bol}/L_{\rm Edd}$} & \colhead{log $M_*$} \\
\colhead{(1)} & \colhead{(2)} & \colhead{(3)} &
\colhead{(4)} & \colhead{(5)} & \colhead{(6)} &
\colhead{(7)} & \colhead{(8)} & \colhead{(9)} 
}
\startdata
  1   &  $-$20.54 &   $-$15.78 &   $-$20.53  &  40.10   &     39.82     &      6.24  &    $-$1.61    &    10.65\tablenotemark{a}  \\  
  2   &  $-$18.49 &   $-$15.38 &   $-$18.43  &  39.92   &     39.60     &      5.48  &    $-$1.01    &     9.80\tablenotemark{a}  \\  
  3   &  $-$19.68 &   $-$18.12 &   $-$19.38  &  41.18   &     41.02     &      6.14  &    $-$0.58    &    10.04\tablenotemark{c}    
\enddata
\tablecomments{
Col. (1): Identification number assigned in this paper.
Col. (2): Total $g$-band absolute magnitude.
Col. (3): AGN $g$-band absolute magnitude, estimated from $L_{\rm
H\alpha^B}$ given in col. (5) and a conversion from $L_{\rm
H\alpha}$ to $M_g$ assuming $f_{\lambda} \propto \lambda^{-1.56}$. %
Col. (4): Host galaxy $g$-band absolute magnitude, obtained by
subtracting the AGN luminosity from the total luminosity.
Col. (5): Luminosity of broad H$\alpha$, in units of \lum.  %
Col. (6): Luminosity of [O\,{\tiny III}] $\lambda5007$, in units of \lum.  %
Col. (7): Virial mass estimate of the BH calculating following GH07 and Dong+12, in units of $M_{\odot}$.  %
Col. (8): Eddington ratio.
Col. (9): Stellar mass of the host galaxy.
{\it (This table is available in its entirety in a machine-readable form in the online
journal. A portion is shown here for guidance regarding its form and content.)}
}
\tablenotetext{a}{Estimated using the color--$M_*/L$ relation in \citet{into13} where color is the 
SDSS $r-i$,  and L is the luminosity of 2MASS Ks-band.}
\tablenotetext{b}{Estimated using the method similar with $a$ but using UKIDSS K-band magnitude.}
\tablenotetext{c}{Derived from MPA--JHU catalog.}
\tablenotetext{d}{Estimated using the scaling relation between WISE 
3.4$\mu$m luminosity and stellar mass provided by \citet{wenxq13}, 
which is calibrated by the stellar masses from the MPA--JHU catalog.}
\label{tab-mbh}
\end{deluxetable}


\begin{deluxetable}{rccccccc}
\tablecolumns{8} \tabletypesize{\scriptsize} \tablewidth{0pc}
\tablecaption{\rosat\ Detections} %
\tablehead{ \colhead{ID}  &
\colhead{Count Rate} & \colhead{$N_{\rm H}$} &
\colhead{$\Gamma$} & \colhead{$f_{\rm 0.5-2.0~keV}$} &
\colhead{log $L_{\rm 0.5-2.0~keV}$}  & \colhead{$L_{2500~\AA}$} &
\colhead{$\aox$}\\
\colhead{(1)} & \colhead{(2)} & \colhead{(3)} &
\colhead{(4)} & \colhead{(5)} & \colhead{(6)} &
\colhead{(7)} & \colhead{(8)} }
\startdata
\multicolumn{8}{c}{\it Our new sample}\\
\hline
    6 &  0.074 $\pm$ 0.014  &  4.40 &  $2.71^{+0.22}_{-0.23}$  &  $-$12.24 $\pm$ 0.08 &  42.31 $\pm$ 0.08  & 27.57  & $-$1.27  \\
  27 &  0.058 $\pm$ 0.014  &  3.10 &  $2.14^{+0.28}_{-0.31}$  &  $-$12.35 $\pm$ 0.11 &  43.94 $\pm$ 0.11   & 28.57  & $-$0.92  \\
  34 &  0.043 $\pm$ 0.014  &  3.50 &  \nodata                 &  $-$12.49 $\pm$ 0.14 &  42.76 $\pm$ 0.14   & 27.96  & $-$1.19  \\
  37 &  0.116 $\pm$ 0.019  &  3.60 &  $2.49^{+0.20}_{-0.22}$  &  $-$12.08 $\pm$ 0.07 &  42.88 $\pm$ 0.07   & 27.92  & $-$1.15  \\
  39 &  0.174 $\pm$ 0.022  &  1.70 &  $2.89^{+1.42}_{-1.33}$  &  $-$12.32 $\pm$ 0.06 &  42.34 $\pm$ 0.06   & 27.63  & $-$1.30  \\
  45 &  0.031 $\pm$ 0.012  &  1.50 &  $1.56^{+0.49}_{-0.67}$  &  $-$12.63 $\pm$ 0.16 &  43.29 $\pm$ 0.16   & 28.16  & $-$0.96  \\
  47 &  0.061 $\pm$ 0.013  &  2.60 &  $2.25^{+0.33}_{-0.34}$  &  $-$12.40 $\pm$ 0.10 &  42.63 $\pm$ 0.10   & 27.64  & $-$1.11  \\
  59 &  0.049 $\pm$ 0.014  &  1.60 &  \nodata                 &  $-$12.67 $\pm$ 0.12 &  42.77 $\pm$ 0.12   & 27.78  & $-$1.12  \\
  75 &  0.005 $\pm$ 0.001  &  2.10 &  $2.31^{+0.29}_{-0.33}$  &  $-$13.59 $\pm$ 0.05 &  40.86 $\pm$ 0.05   & 26.43  & $-$1.33  \\
  82 &  0.040 $\pm$ 0.012  &  1.80 &  \nodata                 &  $-$12.72 $\pm$ 0.12 &  42.92 $\pm$ 0.12   & 28.00  & $-$1.14  \\
  87 &  0.072 $\pm$ 0.015  &  1.70 &  \nodata                 &  $-$12.49 $\pm$ 0.09 &  43.68 $\pm$ 0.09   & 28.60  & $-$1.06  \\
  96 &  0.014 $\pm$ 0.002  &  1.40 &  1.73 $\pm$ 0.16         &  $-$13.05 $\pm$ 0.05 &  41.84 $\pm$ 0.05   & 26.87  & $-$1.06  \\
  99 &  0.058 $\pm$ 0.000  &  1.60 &  \nodata                 &  $-$12.60 $\pm$ 0.00 &  43.29 $\pm$ 0.00   & 28.08  & $-$1.02  \\
 100 &  0.176 $\pm$ 0.023  &  2.30 &  $1.71^{+0.15}_{-0.17}$  &  $-$11.84 $\pm$ 0.06 &  43.27 $\pm$ 0.06   & 27.88  & $-$0.89  \\
 103 &  0.072 $\pm$ 0.014  &  1.40 &  $2.94^{+0.23}_{-0.28}$  &  $-$12.79 $\pm$ 0.09 &  43.31 $\pm$ 0.09   & 28.13  & $-$1.11  \\
 104 &  0.051 $\pm$ 0.013  &  1.20 &  \nodata                 &  $-$12.73 $\pm$ 0.11 &  43.45 $\pm$ 0.11   & 28.41  & $-$1.08  \\
 110 &  0.007 $\pm$ 0.002  &  2.60 &  2.39 $\pm$ 0.35         &  $-$13.39 $\pm$ 0.10 &  42.63 $\pm$ 0.10   & 28.35  & $-$1.38  \\
 117 &  0.060 $\pm$ 0.013  &  1.30 &  2.04 $\pm$ 0.16         &  $-$12.52 $\pm$ 0.09 &  41.71 $\pm$ 0.09   & 27.06  & $-$1.22  \\
 118 &  0.022 $\pm$ 0.009  &  1.10 &  \nodata                 &  $-$13.12 $\pm$ 0.18 &  42.66 $\pm$ 0.18   & 27.84  & $-$1.18  \\
 121 &  0.027 $\pm$ 0.011  &  1.10 &  \nodata                 &  $-$13.04 $\pm$ 0.17 &  41.93 $\pm$ 0.17   & 27.51  & $-$1.34  \\
 123 &  0.698 $\pm$ 0.038  &  1.20 &  $2.83^{+0.79}_{-0.76}$  &  $-$11.81 $\pm$ 0.02 &  43.55 $\pm$ 0.02   & 28.37  & $-$1.11  \\
 124 &  0.098 $\pm$ 0.016  &  1.10 &  $1.57^{+0.20}_{-0.22}$  &  $-$12.19 $\pm$ 0.07 &  43.48 $\pm$ 0.07   & 28.10  & $-$0.87  \\
 125 &  0.052 $\pm$ 0.012  &  1.10 &  \nodata                 &  $-$12.74 $\pm$ 0.10 &  42.59 $\pm$ 0.10   & 28.19  & $-$1.35  \\
 126 &  0.062 $\pm$ 0.013  &  1.20 &  $1.26^{+0.20}_{-0.21}$  &  $-$12.30 $\pm$ 0.09 &  42.79 $\pm$ 0.09   & 28.15  & $-$1.13  \\
 130 &  0.297 $\pm$ 0.023  &  1.20 &  $2.43^{+1.02}_{-1.01}$  &  $-$12.00 $\pm$ 0.03 &  42.74 $\pm$ 0.03   & 27.69  & $-$1.11  \\
 132 &  0.053 $\pm$ 0.012  &  0.90 &  \nodata                 &  $-$12.79 $\pm$ 0.10 &  43.36 $\pm$ 0.10   & 28.26  & $-$1.06  \\
 135 &  0.052 $\pm$ 0.000  &  1.10 &  $2.88^{+0.77}_{-0.75}$  &  $-$12.99 $\pm$ 0.00 &  42.03 $\pm$ 0.00   & 28.01  & $-$1.56  \\
 138 &  0.083 $\pm$ 0.023  &  2.00 &  \nodata                 &  $-$12.38 $\pm$ 0.12 &  42.83 $\pm$ 0.12   & 28.00  & $-$1.18  \\
 144 &  0.029 $\pm$ 0.010  &  1.40 &  $1.93^{+0.26}_{-0.27}$  &  $-$12.78 $\pm$ 0.15 &  43.06 $\pm$ 0.15   & 28.05  & $-$1.05  \\
 150 &  0.028 $\pm$ 0.004  &  1.60 &  1.78 $\pm$ 0.18         &  $-$12.73 $\pm$ 0.06 &  42.97 $\pm$ 0.06   & 28.05  & $-$1.07  \\
 155 &  0.098 $\pm$ 0.015  &  1.00 &  \nodata                 &  $-$12.49 $\pm$ 0.07 &  43.25 $\pm$ 0.07   & 28.14  & $-$1.06  \\
 165 &  0.036 $\pm$ 0.014  &  1.90 &  \nodata                 &  $-$12.75 $\pm$ 0.17 &  43.30 $\pm$ 0.17   & 28.34  & $-$1.11  \\
 166 &  0.095 $\pm$ 0.005  &  2.40 &  2.05 $\pm$ 0.11         &  $-$12.18 $\pm$ 0.02 &  42.74 $\pm$ 0.02   & 27.35  & $-$0.93  \\
 176 &  0.023 $\pm$ 0.010  &  3.80 &  \nodata                 &  $-$12.73 $\pm$ 0.18 &  42.45 $\pm$ 0.18   & 27.49  & $-$1.13  \\
 177 &  0.116 $\pm$ 0.022  &  3.90 &  $2.37^{+0.29}_{-0.33}$  &  $-$12.03 $\pm$ 0.08 &  42.55 $\pm$ 0.08   & 27.32  & $-$1.04  \\
 186 &  0.047 $\pm$ 0.012  &  5.50 &  \nodata                 &  $-$12.32 $\pm$ 0.11 &  42.38 $\pm$ 0.11   & 27.62  & $-$1.21  \\
 190 &  0.015 $\pm$ 0.001  &  3.50 &  $1.27^{+0.18}_{-0.23}$  &  $-$12.78 $\pm$ 0.04 &  41.56 $\pm$ 0.04   & 26.35  & $-$0.92  \\
 192 &  0.111 $\pm$ 0.015  &  1.70 &  $1.65^{+0.17}_{-0.18}$  &  $-$12.09 $\pm$ 0.06 &  43.46 $\pm$ 0.06   & 28.15  & $-$0.91  \\
\hline
\multicolumn{8}{c}{\it Dong+12} \\
\hline
  10 &  0.050 $\pm$ 0.013  & 3.9 & $2.53^{+0.27}_{-0.29}$  &  $-$12.42$\pm$0.11  &  42.76$\pm$0.11  & 27.86 &  $-$1.18  \\
  15 &  0.030 $\pm$ 0.014  & 5.3 & $2.56^{+0.37}_{-0.44}$  &  $-$12.55$\pm$0.20  &  43.31$\pm$0.20  & 27.94 &  $-$0.99  \\
  18 &  0.024 $\pm$ 0.004  & 3.6 & $1.55^{+0.16}_{-0.19}$  &  $-$12.61$\pm$0.13  &  42.48$\pm$0.13  & 27.15 &  $-$0.90  \\
  21 &  0.067 $\pm$ 0.019  & 7.2 & $2.42^{+0.17}_{-0.16}$  &  $-$12.11$\pm$0.19  &  42.57$\pm$0.19  & 27.82 &  $-$1.23  \\
  24 &  0.026 $\pm$ 0.011  & 8.1 & \nodata                 &  $-$12.49$\pm$0.16  &  43.16$\pm$0.16  & 28.30 &  $-$1.16  \\
  27 &  0.026 $\pm$ 0.010  & 6.2 & \nodata                 &  $-$12.54$\pm$0.08  &  42.21$\pm$0.08  & 27.55 &  $-$1.25  \\
  67 &  0.145 $\pm$ 0.025  & 2.8 & $2.91^{+0.31}_{-0.32}$  &  $-$12.20$\pm$0.13  &  42.83$\pm$0.13  & 27.43 &  $-$1.04  \\
  68 &  0.046 $\pm$ 0.014  & 3.0 & $2.39^{+0.17}_{-0.29}$  &  $-$12.52$\pm$0.11  &  43.19$\pm$0.11  & 28.03 &  $-$1.05  \\
  74 &  0.061 $\pm$ 0.015  & 4.9 & $3.83^{+0.27}_{-0.28}$  &  $-$12.55$\pm$0.08  &  42.75$\pm$0.08  & 27.84 &  $-$1.36  \\
  78 &  0.105 $\pm$ 0.019  & 4.3 & $3.06^{+0.63}_{-0.45}$  &  $-$12.18$\pm$0.12  &  43.60$\pm$0.12  & 28.11 &  $-$1.01  \\
  79 &  0.053 $\pm$ 0.014  & 3.7 & $3.96^{+0.34}_{-0.27}$  &  $-$12.86$\pm$0.08  &  43.98$\pm$0.08  & 28.76 &  $-$1.23  \\
  87 &  0.113 $\pm$ 0.022  & 1.7 & $3.24^{+1.27}_{-0.87}$  &  $-$12.66$\pm$0.16  &  42.26$\pm$0.16  & 27.69 &  $-$1.40  \\
  98 &  0.034 $\pm$ 0.013  & 1.6 & \nodata                 &  $-$12.83$\pm$0.07  &  42.23$\pm$0.07  & 27.95 &  $-$1.39  \\
  99 &  0.107 $\pm$ 0.016  & 0.9 & \nodata                 &  $-$12.48$\pm$0.13  &  44.03$\pm$0.13  & 28.40 &  $-$0.84  \\
 101 &  0.072 $\pm$ 0.000  & 0.7 & \nodata                 &  $-$12.60$\pm$0.20  &  42.17$\pm$0.20  & 26.95 &  $-$1.01  \\
 102 &  0.040 $\pm$ 0.012  & 1.3 & \nodata                 &  $-$12.82$\pm$0.14  &  42.68$\pm$0.14  & 27.89 &  $-$1.19  \\
 103 &  0.017 $\pm$ 0.008  & 2.5 & \nodata                 &  $-$13.01$\pm$0.14  &  43.09$\pm$0.14  & 28.24 &  $-$1.16  \\
 105 &  0.033 $\pm$ 0.011  & 3.5 & \nodata                 &  $-$12.60$\pm$0.06  &  42.76$\pm$0.06  & 27.53 &  $-$1.03  \\
 110 &  0.019 $\pm$ 0.002  & 3.0 & $2.61^{+0.31}_{-0.30}$  &  $-$12.95$\pm$0.12  &  42.47$\pm$0.12  & 27.18 &  $-$1.03  \\
 112 &  0.029 $\pm$ 0.009  & 1.3 & \nodata                 &  $-$12.96$\pm$0.10  &  39.90$\pm$0.10  & 25.45 &  $-$1.34  \\
 115 &  0.127 $\pm$ 0.017  & 0.7 & $2.53^{+0.15}_{-0.14}$  &  $-$12.55$\pm$0.05  &  42.54$\pm$0.05  & 27.01 &  $-$0.93  \\
 117 &  0.002 $\pm$ 0.001  & 4.3 & 2.28 $\pm$ 0.05         &  $-$13.75$\pm$0.06  &  41.79$\pm$0.06  & 28.04 &  $-$1.58  \\
 123 &  0.046 $\pm$ 0.013  & 3.0 & $2.57^{+0.27}_{-0.22}$  &  $-$12.56$\pm$0.15  &  43.96$\pm$0.15  & 28.85 &  $-$1.07  \\
 125 &  0.067 $\pm$ 0.016  & 1.4 & $2.37^{+0.42}_{-0.36}$  &  $-$12.57$\pm$0.12  &  43.23$\pm$0.12  & 28.41 &  $-$1.18  \\
 135 &  0.005 $\pm$ 0.002  & 0.8 & $1.51^{+0.05}_{-0.06}$  &  $-$13.50$\pm$0.08  &  41.57$\pm$0.08  & 27.09 &  $-$1.22  \\
 138 &  0.099 $\pm$ 0.000  & 1.9 & $1.46^{+0.33}_{-0.28}$  &  $-$12.52$\pm$0.15  &  43.62$\pm$0.15  & 27.76 &  $-$0.84  \\
 149 &  0.032 $\pm$ 0.003  & 1.1 & 1.21 $\pm$ 0.13         &  $-$13.12$\pm$0.12  &  42.43$\pm$0.12  & 27.71 &  $-$1.26  \\
 153 &  0.115 $\pm$ 0.000  & 2.4 & $1.96^{+0.30}_{-0.41}$  &  $-$12.19$\pm$0.09  &  43.03$\pm$0.09  & 27.89 &  $-$1.07  \\
 163 &  0.176 $\pm$ 0.021  & 1.1 & $2.52^{+0.52}_{-0.44}$  &  $-$12.29$\pm$0.17  &  42.51$\pm$0.17  & 27.48 &  $-$1.13  \\
 164 &  0.226 $\pm$ 0.029  & 2.0 & $2.10^{+0.11}_{-0.12}$  &  $-$11.86$\pm$0.03  &  42.27$\pm$0.03  & 27.77 &  $-$1.28  \\
 169 &  0.041 $\pm$ 0.014  & 1.9 & \nodata                 &  $-$12.70$\pm$0.13  &  42.40$\pm$0.13  & 27.56 &  $-$1.18  \\
 174 &  0.044 $\pm$ 0.012  & 1.2 & \nodata                 &  $-$12.72$\pm$0.06  &  40.95$\pm$0.06  & 26.84 &  $-$1.44  \\
 175 &  0.101 $\pm$ 0.019  & 1.6 & 1.56 $\pm$ 0.16         &  $-$12.12$\pm$0.05  &  43.01$\pm$0.05  & 27.60 &  $-$0.87  \\
 183 &  0.030 $\pm$ 0.011  & 1.0 & \nodata                 &  $-$13.01$\pm$0.09  &  42.46$\pm$0.09  & 27.37 &  $-$1.08  \\
 184 &  0.045 $\pm$ 0.012  & 1.4 & \nodata                 &  $-$12.75$\pm$0.13  &  43.50$\pm$0.13  & 28.54 &  $-$1.11  \\
 196 &  0.101 $\pm$ 0.022  & 2.0 & $1.44^{+0.30}_{-0.32}$  &  $-$12.06$\pm$0.14  &  42.23$\pm$0.14  & 27.33 &  $-$1.06  \\
 203 &  0.021 $\pm$ 0.008  & 1.2 & $2.06^{+0.77}_{-0.96}$  &  $-$13.01$\pm$0.09  &  42.03$\pm$0.09  & 27.67 &  $-$1.33  \\
 212 &  0.396 $\pm$ 0.027  & 0.9 & \nodata                 &  $-$11.91$\pm$0.07  &  43.67$\pm$0.07  & 27.98 &  $-$0.85  \\
 214 &  0.032 $\pm$ 0.010  & 1.7 & \nodata                 &  $-$12.81$\pm$0.13  &  42.52$\pm$0.13  & 27.98 &  $-$1.28  \\
 217 &  0.125 $\pm$ 0.016  & 1.8 & $2.36^{+0.31}_{-0.33}$  &  $-$12.23$\pm$0.16  &  42.77$\pm$0.16  & 27.69 &  $-$1.08  \\
 221 &  0.121 $\pm$ 0.015  & 2.0 & $2.24^{+0.17}_{-0.16}$  &  $-$12.17$\pm$0.06  &  43.30$\pm$0.06  & 28.12 &  $-$1.03  \\
 229 &  0.041 $\pm$ 0.008  & 1.6 & \nodata                 &  $-$12.75$\pm$0.07  &  42.97$\pm$0.07  & 27.88 &  $-$1.07  \\
 230 &  0.027 $\pm$ 0.008  & 1.8 & \nodata                 &  $-$12.89$\pm$0.05  &  42.81$\pm$0.05  & 27.98 &  $-$1.17  \\
 232 &  0.043 $\pm$ 0.014  & 2.8 & $2.28^{+0.37}_{-0.28}$  &  $-$12.54$\pm$0.08  &  42.95$\pm$0.08  & 27.99 &  $-$1.12  \\
 235 &  0.055 $\pm$ 0.011  & 1.4 & \nodata                 &  $-$11.90$\pm$0.09  &  43.62$\pm$0.09  & 27.64 &  $-$0.73  \\
 237 &  0.098 $\pm$ 0.016  & 1.5 & $2.89^{+0.29}_{-0.24}$  &  $-$12.61$\pm$0.07  &  43.04$\pm$0.07  & 28.18 &  $-$1.23  \\
 239 &  0.047 $\pm$ 0.014  & 2.6 & $2.17^{+0.37}_{-0.41}$  &  $-$12.55$\pm$0.12  &  41.72$\pm$0.12  & 26.70 &  $-$1.12  \\
 244 &  0.052 $\pm$ 0.005  & 1.3 & $1.69^{+0.05}_{-0.06}$  &  $-$12.58$\pm$0.02  &  42.76$\pm$0.02  & 27.96 &  $-$1.15  \\
 246 &  0.024 $\pm$ 0.009  & 1.0 & 1.63 $\pm$ 0.15         &  $-$12.83$\pm$0.06  &  42.35$\pm$0.06  & 27.45 &  $-$1.07  \\
 249 &  0.095 $\pm$ 0.013  & 1.4 & $2.20^{+0.52}_{-0.46}$  &  $-$12.36$\pm$0.08  &  43.10$\pm$0.08  & 28.14 &  $-$1.11  \\
 252 &  0.073 $\pm$ 0.012  & 1.4 & $2.49^{+0.16}_{-0.18}$  &  $-$12.59$\pm$0.12  &  42.72$\pm$0.12  & 27.95 &  $-$1.22  \\
 253 &  0.107 $\pm$ 0.013  & 1.7 & $1.96^{+0.37}_{-0.42}$  &  $-$12.18$\pm$0.10  &  42.95$\pm$0.10  & 28.04 &  $-$1.10  \\
 256 &  0.103 $\pm$ 0.018  & 1.5 & $2.87^{+0.39}_{-0.38}$  &  $-$12.58$\pm$0.14  &  42.67$\pm$0.14  & 27.88 &  $-$1.26  \\
 259 &  0.038 $\pm$ 0.008  & 1.3 & $2.61^{+0.46}_{-0.45}$  &  $-$12.95$\pm$0.18  &  42.93$\pm$0.18  & 27.73 &  $-$1.06  \\
 263 &  0.185 $\pm$ 0.028  & 2.4 & $3.37^{+0.40}_{-0.45}$  &  $-$12.34$\pm$0.14  &  42.76$\pm$0.14  & 27.77 &  $-$1.26  \\
 269 &  0.042 $\pm$ 0.011  & 6.6 & 3.36 $\pm$ 0.10         &  $-$12.44$\pm$0.06  &  42.99$\pm$0.06  & 28.02 &  $-$1.26  \\
 271 &  0.895 $\pm$ 0.035  & 1.3 & 2.55 $\pm$ 1.21         &  $-$11.55$\pm$0.00  &  42.89$\pm$0.00  & 27.57 &  $-$1.03  \\
 276 &  0.088 $\pm$ 0.013  & 1.3 & \nodata                 &  $-$12.40$\pm$0.04  &  42.03$\pm$0.04  & 27.10 &  $-$1.12  \\
 284 &  0.017 $\pm$ 0.003  & 2.5 & $2.65^{+0.78}_{-1.06}$  &  $-$13.09$\pm$0.15  &  42.24$\pm$0.15  & 27.77 &  $-$1.36  \\
 286 &  0.017 $\pm$ 0.005  & 2.8 & $1.51^{+0.52}_{-0.61}$  &  $-$12.81$\pm$0.19  &  42.84$\pm$0.19  & 27.68 &  $-$0.95  \\
 287 &  0.022 $\pm$ 0.005  & 3.1 & $2.20^{+0.53}_{-0.52}$  &  $-$12.79$\pm$0.00  &  42.62$\pm$0.00  & 27.73 &  $-$1.14  \\
 289 &  0.034 $\pm$ 0.011  & 5.0 & $3.34^{+1.23}_{-1.18}$  &  $-$12.66$\pm$0.03  &  42.51$\pm$0.03  & 27.98 &  $-$1.43  \\
 297 &  0.021 $\pm$ 0.009  & 4.9 & $3.20^{+0.75}_{-0.73}$  &  $-$12.85$\pm$0.00  &  43.05$\pm$0.00  & 28.27 &  $-$1.30  \\
 301 &  0.045 $\pm$ 0.014  & 2.9 & $2.20^{+0.89}_{-0.88}$  &  $-$12.49$\pm$0.04  &  43.76$\pm$0.04  & 27.79 &  $-$0.70   
\enddata
\tablecomments{
Col. (1): Identification number assigned in this paper and Dong+12 respectively.
Col. (2): \rosat\ count rate (count s$^{-1}$).
Col. (3): Galactic column density $N_{\rm H}$\,(10$^{20}$~cm$^{-2}$).
Col. (4): Photon index of X-ray spectrum.
Col. (5): X-ray flux in the 0.5$-$2.0\,keV in log-scale ($\mathrm{erg~s^{-1}~cm^{-2}}$).
Col. (6): X-ray luminosity in the 0.5$-$2.0\,keV in log-scale (\lum).
Col. (7): Monochromatic luminosity at 2500~\AA\,(erg~s$^{-1}$~Hz$^{-1}$).
Col. (8): Optical-to-X-ray effective spectral index \aox.
}
\label{tab-rosat}
\end{deluxetable}


\begin{deluxetable}{rccc}
\tablecolumns{4} \tabletypesize{\scriptsize} \tablewidth{0pc}
\tablecaption{FIRST Detections} %
\tablehead{ \colhead{ID}  &
\colhead{$S_{\rm 20cm}$} & \colhead{log~$P_{\rm 20cm}$} &
\colhead{log~$R$} \\
\colhead{(1)} & \colhead{(2)} & \colhead{(3)} &
\colhead{(4)} }
\startdata
\multicolumn{4}{c}{\it Our new sample} \\
\hline
 23     &    1.14   $\pm$ 0.16  & 22.80 $\pm$ 0.06 & 1.36  \\
 28     &    0.85   $\pm$ 0.15  & 22.15 $\pm$ 0.07 & 1.22  \\
 55     &    2.86   $\pm$ 0.19  & 22.32 $\pm$ 0.03 & 1.49  \\
 89     &    1.22   $\pm$ 0.16  & 22.89 $\pm$ 0.06 & 1.51  \\
 95     &    16.75  $\pm$ 0.86  & 23.59 $\pm$ 0.02 & 2.49  \\
 123    &    2.72   $\pm$ 0.19  & 22.76 $\pm$ 0.03 & 1.05  \\
 143    &    1.97   $\pm$ 0.17  & 22.20 $\pm$ 0.04 & 1.51  \\
 199    &    2.67   $\pm$ 0.18  & 22.27 $\pm$ 0.03 & 0.80  \\
 201    &    1.79   $\pm$ 0.16  & 23.07 $\pm$ 0.04 & 1.34  \\
\hline
\multicolumn{4}{c}{\it Dong+12} \\
\hline
22  &    1.82 $\pm$ 0.14  &  22.32 $\pm$ 0.03 &  1.57 \\
30  &    3.18 $\pm$ 0.27  &  23.12 $\pm$ 0.04 &  2.03 \\
36  &    4.53 $\pm$ 0.26  &  23.32 $\pm$ 0.03 &  2.06 \\
54  &    2.07 $\pm$ 0.18  &  22.75 $\pm$ 0.04 &  1.43 \\
78  &    1.44 $\pm$ 0.17  &  22.88 $\pm$ 0.05 &  1.44 \\
118 &    1.04 $\pm$ 0.14  &  21.87 $\pm$ 0.06 &  0.84 \\
140 &    2.38 $\pm$ 0.19  &  22.05 $\pm$ 0.03 &  1.57 \\
141 &    2.55 $\pm$ 0.18  &  21.83 $\pm$ 0.03 &  1.58 \\
170 &    1.03 $\pm$ 0.16  &  21.95 $\pm$ 0.07 &  1.58 \\
181 &    1.29 $\pm$ 0.16  &  22.32 $\pm$ 0.06 &  0.91 \\
206 &    2.33 $\pm$ 0.18  &  21.39 $\pm$ 0.03 &  1.49 \\
208 &    8.81 $\pm$ 0.46  &  23.20 $\pm$ 0.02 &  2.21 \\
222 &    2.69 $\pm$ 0.20  &  21.58 $\pm$ 0.03 &  1.15 \\
257 &    2.52 $\pm$ 0.19  &  22.56 $\pm$ 0.03 &  1.50 \\
273 &    1.43 $\pm$ 0.16  &  22.53 $\pm$ 0.05 &  1.09 \\
277 &    0.68 $\pm$ 0.14  &  23.02 $\pm$ 0.09 &  1.36 \\
305 &    3.78 $\pm$ 0.23  &  22.79 $\pm$ 0.03 &  1.98 
\enddata
\tablecomments{
Col. (1): Identification number assigned in this paper and Dong+12 respectively.
Col. (2): Flux density at 20 cm from the FIRST (mJy). Uncertainties include the 5\% systematic uncertainty recommended by \citet{white97}.
Col. (3): Corresponding radio power (W~$\rm Hz^{-1}$) at 20~cm.
Col. (4): Radio loudness in log-scale. \rkel\ $\equiv f_{\rm 6cm}/f_{\rm 4400 
\angstrom}$  assuming a spectral index of $\alpha_{r}\,= 0.46$
\citep[radio;][]{ulvestad01} and $\alpha_{o}\,= 0.44$ \citep[optical;][]
{vandenberk01}, where $f_{\nu} \propto {\nu^{-\alpha_{o}}}$.
}
\label{tab-first}
\end{deluxetable}

\clearpage
\end{document}